%% file: main.tex
\documentclass[journal,twoside,web]{ieeecolor}
\usepackage{generic}
\usepackage{cite}
\usepackage{amsmath,amssymb,amsfonts}
\usepackage{algorithmic}
\usepackage{graphicx}
\usepackage{textcomp}
\usepackage{subcaption}
\usepackage{url}

\newtheorem{proposition}{Proposition}
\newtheorem{Lemma}{Lemma}
\newtheorem{Theorem}{Theorem}
\newtheorem{corollary}{Corollary}
\newtheorem{Definition}{Definition}
\newtheorem{assumption}{Assumption}

\def\BibTeX{{\rm B\kern-.05em{\sc i\kern-.025em b}\kern-.08em
    T\kern-.1667em\lower.7ex\hbox{E}\kern-.125emX}}
\begin{document}
\title{Safe Control for Nonlinear Systems under Faults and Attacks via Control Barrier Functions}
\author{Hongchao Zhang,~\IEEEmembership{Student Member,~IEEE,} Zhouchi Li,~\IEEEmembership{Student Member,~IEEE,} and Andrew Clark,~\IEEEmembership{Senior Member,~IEEE}
\thanks{H. Zhang, and A. Clark are with the Electrical and Systems Engineering Department, McKelvey School of Engineering, Washington University in St. Louis, St. Louis, MO 63130 USA (e-mail:\{hongchao,andrewclark\}@wustl.edu) }
\thanks{Z. Li is with Black Sesame Technologies Inc, San Jose, CA 95131 USA (e-mail:  lizhouchi@gmail.com).}
\thanks{This work was supported by National Science Foundation grants CNS-1941670, CMMI-2418806, 
and Air Force Office of Scientific Research grant FA9550-22-1-0054.}
}

\maketitle

\begin{abstract}
Safety is one of the most important properties of control systems. Sensor faults and attacks and actuator failures may cause errors in the sensor measurements and system dynamics, which leads to erroneous control inputs and hence safety violations. In this paper, we improve the robustness against sensor faults and actuator failures by proposing a class of Fault-Tolerant Control Barrier Functions (FT-CBFs) for nonlinear systems. Our approach maintains a set of state estimators according to fault patterns and incorporates CBF-based constraints to ensure safety under sensor faults. We then propose a framework for joint safety and stability by integrating FT-CBFs with Control Lyapunov Functions. By utilizing redundancy, we proposed High order CBF-based approach to ensure safety when actuator failures occur. We propose a sum-of-squares (SOS) based approach to verify the feasibility of FT-CBFs for both sensor faults and actuator failures. We evaluate our approach via two case studies, namely, a wheeled mobile robot (WMR) system in the presence of a sensor attack and a Boeing 747 lateral control system under actuator failures.
\end{abstract}


\input{sections/Intro}

\input{sections/RelatedWork}

\input{sections/Preliminaries}
\input{sections/SensorFault}

\input{sections/FeasibilityVerification}

\input{sections/SF_joint}

\input{sections/ActuatorFailure}

\input{sections/CaseStudy}

\section{Conclusion}
\label{sec:conclusion}
This paper proposed a new class of SCBFs with high relative degree for safety and stability of control systems under sensor faults and attacks. 
Our approach maintains a set of state estimators, excludes outlier estimates and ensures safety with a CBF-based approach. 
We then constructed an SCBF with high order degree for each state estimator, which guaranteed safety provided that a linear constraint on the control input was satisfied at each time step. We proposed a scheme for using additional state estimators to resolve conflicts between these constraints, and derived a scheme to verify the feasibility of SCBFs. We then showed how to compose our proposed HOSCBFs with CLFs to provide joint guarantees on safety and stability of a desired goal set under sensor faults and attacks. 
We proposed HOCBF-based approach to ensure safety of systems under all possible actuator failures and proposed an SOS-based scheme to verify the existence of control inputs satisfying HOCBF constraints. 
The proposed approach against sensor faults was validated on a wheeled mobile robot and our approach against actuator failures was validated on a Boeing 747 lateral control system. Future work in this area will include attacks that jointly affect sensors and actuators. 

\bibliographystyle{ieeetr}
\bibliography{ref} 

\begin{IEEEbiography}[{\includegraphics[width=1in,height=1.25in,clip,keepaspectratio]{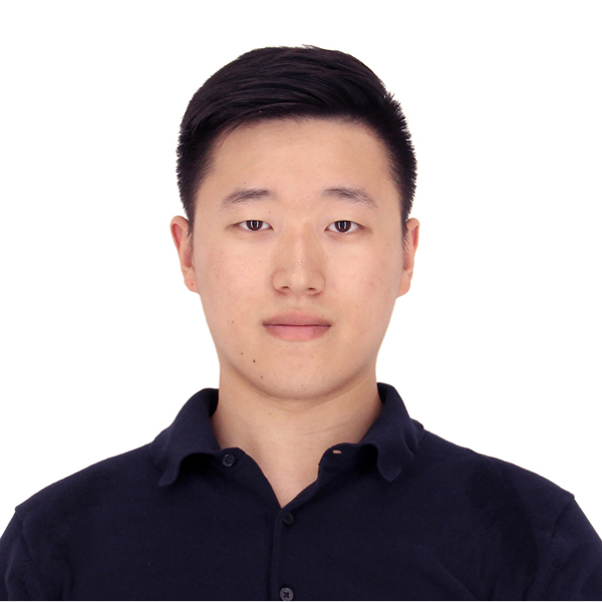}}]{Hongchao Zhang}
received the B.Eng. degree  in Automation Engineering from the Department of Automation Engineering, Nanjing University of Aeronautics and Astronautics, Nanjing, China, in 2018 and the M.Sc. degree in Electrical and Computer Engineering from the Department of Electrical and Computer Engineering, Worcester Polytechnic Institute (WPI) in 2020. 
He is currently working toward the Ph.D. degree in Electrical Engineering with the Department of Electrical and Systems Engineering, Washington University in St. Louis, St. Louis, MO, USA. 
His current research interests include control and security of cyber-physical systems and safe learning-based control. 
Mr Zhang was the recipient of the 2023 General
Motors AutoDriving Security Award at the inaugural ISOC Symposium on Vehicle Security and Privacy at the Network and Distributed System Security Symposium (NDSS). 
\end{IEEEbiography}

\begin{IEEEbiography}[{\includegraphics[width=1in,height=1.25in,clip,keepaspectratio]{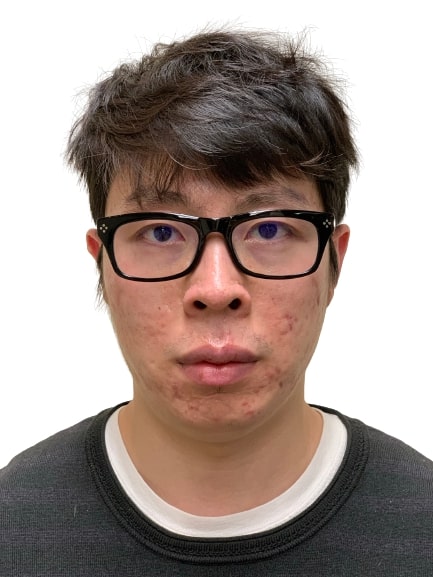}}]{Zhouchi Li}
received the B.Eng. degree in Electronic Science and Technology from the Department of Electronic Science and Technology, Huazhong University of Science and Technology, Wuhan, China, in 2013 and the M.Sc. degree in Electrical and Computer Engineering from the Department of Electrical and Computer Engineering, Worcester Polytechnic Institute in 2016. He received the Ph.D. degree in Electrical Engineering from the Secure Cyber-Physical Systems Lab, Department of Electrical and Computer Engineering,
at Worcester Polytechnic Institute in 2023. He received the 2023 General Motors AutoDriving Security Award at the inaugural ISOC Symposium on Vehicle Security and Privacy at the Network and Distributed System Security Symposium (NDSS). His research interests include control and security of cyber-physical systems.
\end{IEEEbiography}

\begin{IEEEbiography}[{\includegraphics[width=1in,height=1.25in,clip,keepaspectratio]{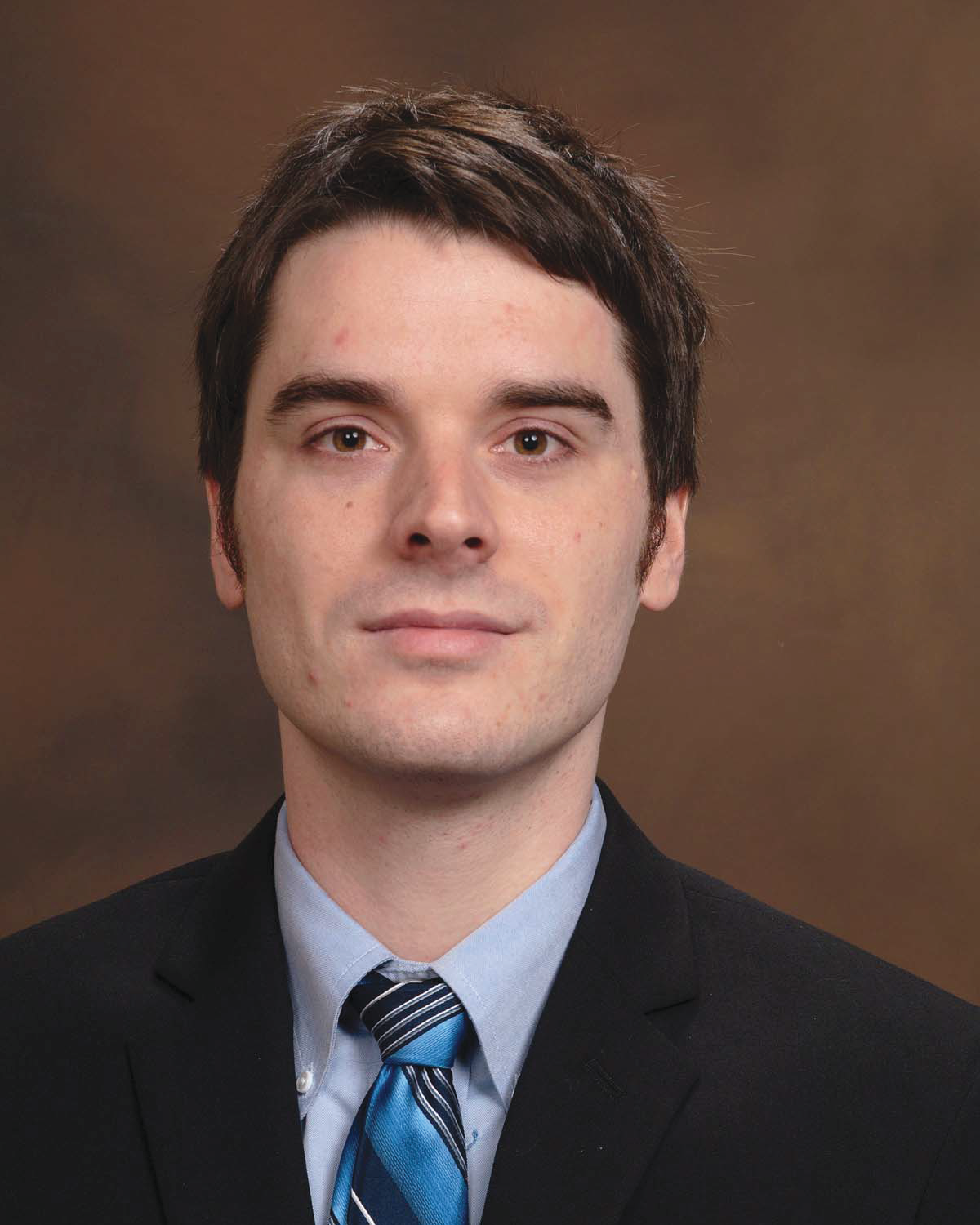}}]{Andrew Clark}
is an Associate Professor in the Department of Electrical and Systems Engineering at Washington University in St. Louis. He received the B.S.E. degree in Electrical Engineering and the M.S. degree in Mathematics from the University of Michigan Ann Arbor in 2007 and 2008, respectively. He received the Ph.D. degree in Electrical Engineering from the Network Security Lab (NSL), Department of Electrical Engineering, at the University of Washington- Seattle in 2014. He is author or co-author of the IEEE/IFIP William C. Carter award-winning paper (2010), the WiOpt Best Paper (2012), the WiOpt Student Best Paper (2014), and the GameSec Outstanding Paper (2018), and was a finalist for the IEEE CDC 2012 Best Student Paper Award and the ACM ICCPS Best Paper Award (2016, 2018, 2020). He received an NSF CAREER award in 2020 and an AFOSR YIP award in 2022. His research interests include control and security of complex networks, safety of autonomous systems, submodular optimization, and control theoretic modeling of network security threats.
\end{IEEEbiography}

\end{document}

%% file: sections/Intro.tex
\section{Introduction}
\label{sec:intro}

A control system is safe if it remains within a predetermined safe region for all time \cite{ames2019control}. 
In applications including medicine, transportation and energy, safety violations can cause catastrophic economic damage and loss of human life \cite{dhscps}.

Approaches verifiable safe control systems include Hamilton–Jacobi–Isaacs (HJI) equation~\cite{tomlin1998conflict}, barrier certificates~\cite{prajna2007framework}, and Control Barrier Functions (CBFs)~\cite{ames2014control}. 
Among those methods, CBFs have the advantage that they can be readily integrated into existing control policies by adding linear constraints on the control input. 
A CBF is a function of the system state that goes to zero as the system approaches the unsafe region. Thus, ensuring safety is equivalent to ensuring that the CBF stays non-negative. 
CBF-based approaches show great promise in safety-critical systems. However, they rely on sensor measurements, which sensor faults, attacks, and actuator failures may compromise. 

Sensor faults and malicious attacks provide inaccurate, arbitrary readings. These inaccurate measurements bias estimates of the system state, leading to erroneous control signals that drive the true system state to an unsafe operating point. 
When actuator failures occur, actuators lose effectiveness and hence render the control system unable to ensure safety or stability \cite{tao2002adaptive}. 
Sensor faults, attacks and actuator failures can cause arbitrary errors in the sensor measurements and system dynamics, which is challenging for existing CBF-based approaches such as \cite{clark2021control} that assume that noises and disturbances are either bounded or come from a known probability distribution. 




Countermeasures such as Fault-Tolerant Control (FTC) \cite{amin2019review,jiang2012fault} have been proposed to accommodate faults, attacks and failures. Existing FTC approaches focus on maintaining performance and do not provide provable safety guarantees. 
Countermeasures incorporating disturbance observer-based CBF are proposed to ensure robust safety of systems with model uncertainties~\cite{wang2022observer,dacs2022robust} and model-free safe reinforcement learning~\cite{cheng2023safe}.
With the growing attention on faults and attacks, safety guarantees on systems under faulty components or adversarial environments have become an active research area.

\begin{figure*}[t!]
\centering
\begin{subfigure}{.41\textwidth}
    \includegraphics[width=\textwidth]{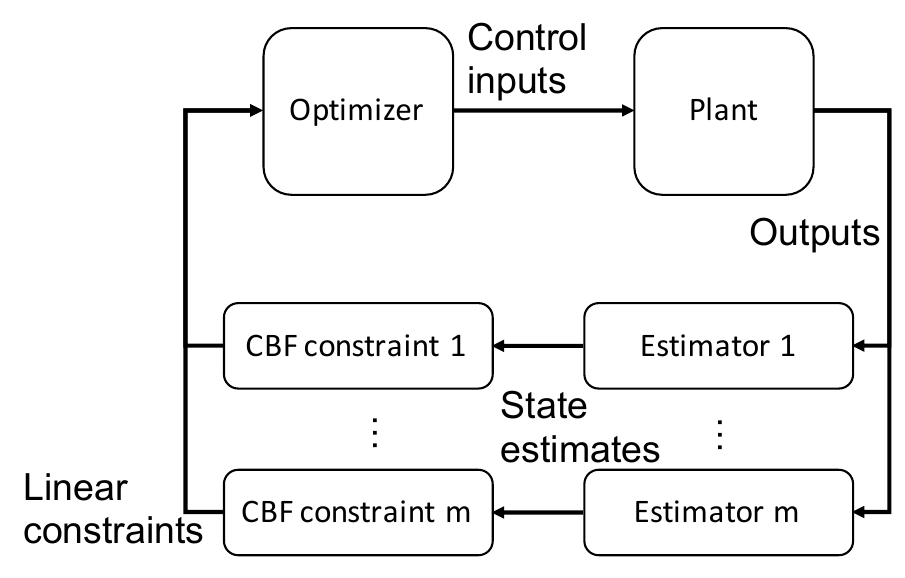}
    \caption{Scheme for system under sensor faults and attacks}
    \label{fig:PM_SF}
\end{subfigure}
\begin{subfigure}{.39\textwidth}
    \includegraphics[width=\textwidth]{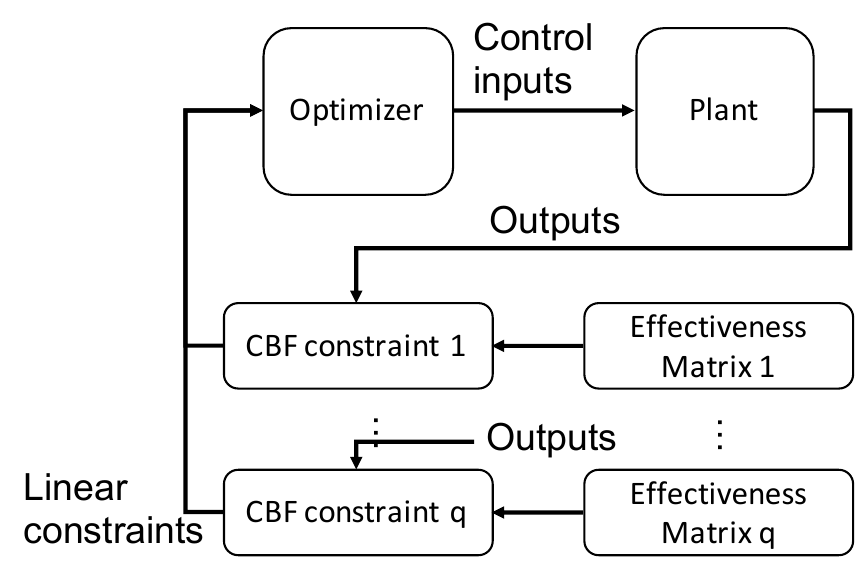}
    \caption{Scheme for system under actuator failures. }
    \label{fig:PM_AF}
\end{subfigure}
\caption{Schematic illustration of our proposed approach. Both schemes ensure safety via CBFs in a passive manner. }
\label{fig:PM_flow}
\end{figure*}

In this paper, we propose safe control algorithms for nonlinear systems under sensor and actuator faults. 
To ensure safety of nonlinear systems under sensor faults and attacks, we propose a class of CBFs, which is shown in Figure \ref{fig:PM_flow}(\subref{fig:PM_SF}) and constructed as follows. We maintain a set of state estimators, each omitting a set of sensors associated with one fault pattern and then use CBF constraints to ensure that each of the estimated states remains within the safe region. 
The intuition of our approach is that one can simply ignore the faulty measurements by omitting the sensors of each fault pattern and construct CBFs for each estimate. However, it may be infeasible to satisfy all CBF constraints using a single control input when faults occur and the state estimates deviate due to the fault. To resolve conflicts, we propose a threshold-based method to exclude outlier estimates and relax the corresponding CBF constraints. 

To ensure safety of nonlinear systems under actuator failures, we propose a class of CBFs, which is shown in Figure \ref{fig:PM_flow}(\subref{fig:PM_AF}) and constructed as follows. We maintain a set of effectiveness matrices according to failure patterns and then use CBFs to ensure the system with each actuator failure remains within the safe region. The basic idea is to find a single control input that satisfies all CBF constraints constructed for each failure pattern. 
We propose a sum-of-squares (SOS) based scheme to verify the feasibility of CBFs for sensor faults/attacks and actuator failures, respectively. 

We make the following specific contributions:
\begin{itemize}
    \item We propose High-order Stochastic CBFs (HOSCBF) for the system with high relative degree and propose FT-SCBFs with high order degree to ensure finite time safety when sensor faults occur. We propose an SOS-based scheme to verify the feasibility of constraints of FT-SCBFs with high relative degree. 
    \item We compose HOSCBFs with Control Lyapunov Functions (CLFs) to provide joint guarantees on safety and stability under sensor faults. 
    \item We formulate an HOCBF approach to ensure safety when actuator failures occur and propose an SOS-based scheme to verify the feasibility of CBF constraints throughout the safe region. 
    \item We evaluate our approach via two case studies. The proposed HOSCBF-CLF ensures safety and convergence of a wheeled mobile robot (WMR) system in the presence of a sensor attack. The proposed HOCBF-based method ensures safety of a Boeing 747 lateral control system under actuator failures. 
\end{itemize}

The remainder of this paper is organized as follows. Section \ref{sec:related} presents the related work. Section \ref{sec:background} presents background and preliminaries. Section \ref{sec:PM_sensorfault} proposes a HOSCBF-based control policy for systems under sensor faults and attacks as well as a scheme to verify the feasibility of HOSCBFs. Section \ref{sec:joint-safety-reach} proposes a framework for joint safety and stability via HOSCBF-CLFs. Section \ref{sec:PM_ActFail} proposes an HOCBF-based control policy for systems under actuator failures. Section \ref{sec:simulation} presents our case studies. Section \ref{sec:conclusion} concludes the paper.

%% file: sections/RelatedWork.tex
\section{Related Work}
\label{sec:related}

Fault-tolerant control systems (FTCS) aim to accommodate faults and maintain stability of the system with little or acceptable degradation in performance. See \cite{amin2019review} for an in-depth treatment. FTCS are classified into two main types, namely, active FTCS and passive FTCS \cite{jiang2012fault}. 
In active FTCS, Fault Detection and Isolation (FDI) plays a significant role and has been studied for decades. See \cite{blanke2006diagnosis} for more details. 

Active FTCS against sensor faults and attacks include statistical hypothesis testing for stochastic systems \cite{chen2019distributed}, and unknown input observers for deterministic systems \cite{li2019performance}. Kalman Filter (KF) and Extended Kalman Filter (EKF) are extensively used in FDI applications such as \cite{bardawily2017sensors} for multiple-input and multiple-output systems. 
More recently, data-driven approaches to fault tolerance have shown promise \cite{wang2018data,jung2018residual}. 
While the approach of using Kalman filter residues to identify potential faults is related to our conflict resolution approach, safety of the system under faults and attacks is not addressed.
Sliding-mode control, as one of popular Passive FTCS, is proposed for singularly perturbed systems \cite{wang2020sliding}, switched systems \cite{yang2021sliding}, fuzzy systems, \cite{liu2021extended}, and Markov Jump Systems \cite{yang2018ito,li2014fault}. 
Several of these works aim to guarantee stability in the presence of faults \cite{yang2018fault}, which is related to but distinct from the safety criteria we consider. 

Passive FTCS against actuator failures is proposed due to its advantage of fast response. 
Reliable control for Linear time-invariant (LTI) system under actuator failure is proposed in \cite{zhao1998reliable} and implemented for LTI aircraft model in \cite{yu2015design}. 
Actuator failure compensation control (AFCC) has been employed for LTI system \cite{tao2002adaptive} and nonlinear systems \cite{wang2010adaptive,tong2013fuzzy}. 
Robust adaptive FTC is presented in \cite{li2012robust} for linear systems with time-varying parameter uncertainty, external disturbance and actuator faults. 
However, the aforementioned methods focus on ensuring stability but leave safety guarantees less studied. In this paper, we address the safety of nonlinear systems under sensor faults/attacks or actuator failures. 



Safety verification of control systems is an area of extensive research, with popular methods including finite-state approximations \cite{girard2012controller}, HJI equation \cite{tomlin1998conflict}, barrier certificates \cite{prajna2007framework,jagtap2020formal}, simulation-driven approaches \cite{kikukawa2008consequence,blom2006particle}, and counterexample-guided synthesis \cite{frehse2008counterexample}.
Barrier function-based approaches, which formulate the safety constraint as inequality over the control input, have been proposed to guarantee safety  \cite{ames2016control,clark2021control,usevitch2020strong,xiao2022control,santoyo2019verification,santoyo2021barrier,choi2021robust}. Among these methods, CBFs were proposed in \cite{ames2014control}. CBFs for stochastic systems were investigated in \cite{clark2021control}. High-order CBFs were presented in \cite{xiao2019control,xiao2022control,nguyen2016exponential}.  CBFs for safe reinforcement learning were introduced in \cite{cheng2019end,marvi2021safe,choi2020reinforcement,ma2021model,zhang2021model}. Applications of CBFs to specific domains such as multi-agent systems \cite{wang2017safety}, autonomous vehicles \cite{xu2017correctness}, and unmanned aerial vehicles \cite{squires2018constructive} have also been considered. Recent research \cite{ames2014control,choi2020reinforcement,xue2022reach,xue2023reach} have investigated joint objectives of safety and stability. The work \cite{usevitch2022adversarial} investigates the resilience of adversarial agents in multi-agent systems with high-order CBFs. None of these existing works, however, incorporated the effects of faults and attacks on actuators and sensors.  

Ensuring safety under sensor faults and attacks has attracted growing research attention. 
Barrier certificate based fault-tolerant Linear quadratic Gaussian (LQG) tracking is investigated in \cite{niu2019lqg} for LTI system under sensor fault and false data injection attack and generalized to multiple possible compromised sensor sets in \cite{li2022lqg}. 
Compared with barrier certificate method, CBF-based approaches have more advantages on flexibility. In the preliminary conference version of this work \cite{clark2020control}, we investigated fault-tolerant control of nonlinear systems, in which CBF constraints are constructed and imposed to ensure safety of the system under sensor faults and attacks. However, systems under actuator failures and fault-tolerant control via CBF with high relative degree have drawn less attention. Recent work \cite{garg2023model} proposed a model-free learning framework for an output-based neural fault-detector to detect actuator faults. However, verifying the feasibility of CBFs under faults has not yet been investigated. 
In this paper, we propose FT-CBFs with high relative degree for nonlinear systems under sensor faults and attacks and actuator failures. We also propose a systematical approach to verify the feasibility of CBFs.  

CBF constructions depend on factors including system dynamics and disturbance, which leave it an open problem to verify whether such constraints can always be satisfied. A systematic approach to verify the feasibility of CBFs is proposed in \cite{clark2021verification} to enable broader adoption of CBFs. However, feasibility verification of CBFs in faulty or adversarial environments has not yet been studied.
In this work, we propose an SOS-based scheme to verify the feasibility of constraints of SCBFs with high relative degree and HOCBFs for the system under sensor faults/attacks and actuator failures, respectively.

%% file: sections/Preliminaries.tex
\section{Preliminaries}
\label{sec:background}
In this section, we present the system model and provide background on the EKF and CBFs. 
\subsection{System Model}
\label{subsec:model}
\noindent\textbf{Notations.} For a set $S$, we denote $\mbox{int}(S)$ and $\partial S$ as the interior and boundary of $S$, respectively. For any vector $v$, we let $[v]_{i}$ denote the $i$-th element of $v$. 
We let $\overline{\lambda}(A)$ denote the magnitude of the largest eigenvalue of matrix $A$, noting that this is equal to the largest eigenvalue when $A$ is symmetric and positive definite. When the value of $A$ is clear, we write $\overline{\lambda}$. 

We consider a nonlinear control system with state $x_{t} \in \mathbb{R}^{n}$ and input $u_{t} \in \mathbb{R}^{p}$ at time $t$. The state dynamics and the system output $y_{t} \in \mathbb{R}^{q}$ are described by the stochastic differential equations
\begin{align}
    \label{eq:state-sde}
    dx_{t} &= (f(x_{t})+g(x_{t})u_{t}) \ dt + \sigma_{t} \ dW_{t}\\
    \label{eq:output-sde}
    dy_{t} &= cx_{t} \ dt + \nu_{t} \ dV_{t}
\end{align}
where $f: \mathbb{R}^{n} \rightarrow \mathbb{R}^{n}$ and $g: \mathbb{R}^{n} \rightarrow \mathbb{R}^{n \times p}$ are locally Lipschitz, $\sigma_{t} \in \mathbb{R}^{n \times n}$, $W_{t}$ is an $n$-dimensional Brownian motion, $c \in \mathbb{R}^{q \times n}$,  $\nu_{t} \in \mathbb{R}^{q \times q}$, and $V_{t}$ is a $q$-dimensional Brownian motion. 

The safety conditions of a system are specified in terms of forward invariance of a pre-defined safe region. We define the safe region as follows.  
\begin{Definition}[Safe Region]
The safe region of the system is a set $\mathcal{C} \subseteq \mathbb{R}^{n}$ defined by 
\begin{equation}
\label{eq:safe-region}
\mathcal{C} = \{{x} : h({x}) \geq 0\}, \quad \partial \mathcal{C} = \{{x} : h({x}) = 0\}
\end{equation}
where $h: \mathbb{R}^{n} \rightarrow \mathbb{R}$ is twice-differentiable on $\mathcal{C}$.
\end{Definition}

We assume throughout the paper that $x_{0} \in \mbox{int}(\mathcal{C})$, i.e., the system is initially safe. 
Let $\overline{f}(x,u) = f(x) + g(x)u$.
The uniform detectability property is defined as follows.
\begin{Definition}[Uniform Detectability]
\label{def:detectability}
The pair $[\frac{\partial \overline{f}}{\partial x}(x,u), c]$ is uniformly detectable if there exists a bounded, matrix-valued function $\Theta(x)$ and a real number $\eta > 0$ such that 
$$w^{T}\left(\frac{\partial \overline{f}}{\partial x}(x,u) + \Theta(x)c\right)w \leq -\eta ||w||^{2}$$
for all $w$, $u$, and $x$. 
\end{Definition}


\subsection{Background and Preliminary Results}
\label{subsec:preliminaries}
Let $\hat{x}_{t}$ denote the EKF estimate of $x_{t}$. The EKF for the system described by \eqref{eq:state-sde} and \eqref{eq:output-sde} is defined by $$d\hat{x}_{t} = (f(\hat{x}_{t}) + g(\hat{x}_{t})u_{t}) dt + K_{t}(dy_{t}-c\hat{x}_{t}),$$ where $K_{t} = P_{t}c^{T}R_{t}^{-1}$ and $R_{t} = \nu_{t}\nu_{t}^{T}$. The matrix $P_{t}$ is the positive-definite solution to $$\frac{dP}{dt} = F_{t}P_{t} + P_{t}F_{t}^{T} + Q_{t} - P_{t}c^{T}R_{t}^{-1}cP_{t}$$ where $Q_{t} = \sigma_{t}\sigma_{t}^{T}$ and $F_{t} = \frac{\partial \overline{f}}{\partial x}(\hat{x}_{t},u_{t})$.  
To utilize EKF, we make the following assumptions.
\begin{assumption}
\label{assumption:ekf}
The SDEs (\ref{eq:state-sde}) and (\ref{eq:output-sde}) satisfy the conditions:
\begin{enumerate}
\item There exist constants $\beta_{1}$ and $\beta_{2}$ such that $\mathbf{E}(\sigma_{t}\sigma_{t}^{T}) \geq \beta_{1} I$ and $\mathbf{E}(\nu_{t}\nu_{t}^{T}) \geq \beta_{2} I$ for all $t$. 
\item The pair $[\frac{\partial \overline{f}}{\partial x}(x,u),c]$ is uniformly detectable.
\item Let $\phi$ be defined by $$\overline{f}(x,u)-\overline{f}(\hat{x},u) = \frac{\partial \overline{f}}{\partial x}(x - \hat{x}) + \phi(x,\hat{x},u).$$ Then there exist real numbers $k_{\phi}$ and $\epsilon_{\phi}$ such that $$||\phi(x,\hat{x},u)|| \leq k_{\phi}||x-\hat{x}||_{2}^{2}$$ for all $x$ and $\hat{x}$ satisfying $||x-\hat{x}||_{2} \leq \epsilon_{\phi}$. 
\end{enumerate}
\end{assumption} 

One can obtain $k_{\phi}$ by considering a compact subset  $\kappa\subseteq\mathbb{R}$. For instance, given a function $f$ that is twice differentiable with respect to $x$ in subset $\kappa$, we have $k_{\phi}=\max_{1\leq i \leq n}\sup_{x\in \kappa}\|\nabla^{2} f_i(x,u)\|$, where $\nabla^{2}f$ is the Hessian matrix. The bounds on estimation error $\epsilon_{\phi}$ can be calculated via an integral formula \cite[Chapter 20]{heuser2013lehrbuch}. More details can be found in~\cite{reif2000stochastic}. 
The following result describes the accuracy of the EKF.

\begin{Theorem}[\cite{reif2000stochastic}]
\label{theorem:EKF}
Suppose that the conditions of Assumption \ref{assumption:ekf} hold. Then there exists $\delta > 0$ such that $\sigma_{t}\sigma_{t}^{T} \leq \delta I$ and $\nu_{t}\nu_{t}^{T} \leq \delta I$. For any $0<\epsilon <1$, there exists $\gamma > 0$ such that $$Pr\left(\sup_{t \geq 0}{||x_{t}-\hat{x}_{t}||_{2} \leq \gamma}\right) \geq 1-\epsilon.$$ 
\end{Theorem}

We next provide background and preliminary results on stochastic CBFs. The following theorem provides sufficient conditions for safety of a stochastic system.
\begin{proposition}
[{\cite[Proposition III.5]{wang2021safety}}]
\label{prop:safety-finite-time}
Given a finite time $T$, suppose the mapping $h$ is a continuous function with linear function $kx$ as the class-$\kappa$ function, where $k\geq 0$. Let the control input $u_{t}$ be chosen to satisfy
\begin{equation}
    \label{eq:SCBF-finite}
    \frac{\partial h}{\partial x}(f(x_t)+g(x_t)u_{t})
    + \frac{1}{2}{tr}\left(\sigma_{t}^{T}\frac{\partial^{2}h}{\partial x^{2}}(x_t)\sigma_{t}\right) \geq -kh(x_t).
\end{equation}
Let $\zeta=\sup_{x\in \mathcal{C}}h(x)$ and $x_0\in int(C)$. 
Then we have $$Pr(x_{t} \in int(\mathcal{C}), 0\leq t\leq T) \geq (\frac{h(x_0)}{\zeta})e^{-\zeta T}.$$
\end{proposition}
\begin{Theorem}
\label{theorem:SCBF}
For a system (1)--(2) with safety region defined by (3), define
$$\overline{h}_{\gamma} = \sup_{x,x^{0}}{\{h(x) : ||x-x^{0}||_{2} \leq \gamma \text{ and } h(x^{0})=0\}}, $$
where $\hat{h}(x) := h(x)-\overline{h}_{\gamma}$.
Let $\frac{\partial h}{\partial x}$ denote the derivative $\frac{\partial h}{\partial x}|_{x=\hat{x}}$ for simplicity. 
Let $z_t \in \mathbb{R}^{n}$ represent the estimation error, where $z_t = (x_t-\hat{x}_t)$. Suppose that $u_{t}$ is chosen to satisfy 
\begin{multline}
    \label{eq:SCBF-2}
    \frac{\partial h}{\partial x}(f(\hat{x}_t)+g(\hat{x}_t)u_{t}) - \left\|\frac{\partial h}{\partial x} K_t c\right\|_2 \gamma \\
    + \frac{1}{2}{tr}\left(\nu_{t}^{T}K_{t}^{T}\frac{\partial^{2}h}{\partial x^{2}}(\hat{x}_t)K_{t}\nu_{t}\right) \geq -\hat{h}(\hat{x}_t).
\end{multline}
Then $Pr(x_{t} \in \mathcal{C}, 0\leq t\leq T  | \ ||x_{t}-\hat{x}_{t}||_{2} \leq \gamma) \geq (\frac{\hat{h}(x_0)}{\zeta})e^{-\zeta T}$.
\end{Theorem}
\begin{proof}
We have the estimate $\hat{x}$ yields 

\begin{align*}
    d \hat{x}_t &=\bar{f}\left(\hat{x}_t, u_t\right) d t+K_t\left(c x_t d t+\nu_t d V_t-c \hat{x}_t d t\right) \\ &=\left(\bar{f}\left(\hat{x}_t, u_t\right)+K_t c\left(x_t-\hat{x}_t\right)\right) d t+K_t \nu_t d V_t
\end{align*}
Given $||x_{t}-\hat{x}_{t}||_{2} \leq \gamma$, we have

$$
\frac{\partial \hat{h}}{\partial x} K_t c\left(x_t-\hat{x}_t\right) \geq -\left\|\frac{\partial \hat{h}}{\partial x} K_t c\right\|_2\left\|z_t\right\|_2 \geq - \left\|\frac{\partial \hat{h}}{\partial x} K_t c\right\|_2 \gamma
$$
We then choose $u_t$ to satisfy \eqref{eq:SCBF-2}. Then, we have 
\begin{equation*}
    \begin{aligned}
    \frac{\partial \hat{h}}{\partial x} \left(f\left(\hat{x}_t\right)+g\left(\hat{x}_t\right) u_t +K_t c\left(x_t-\hat{x}_t\right)\right) & + \\
    \frac{1}{2} \operatorname{tr}\left(\nu_t^T K_t^T\left(\frac{\partial^2 \hat{h}}{\partial x^2}\right) K_t \nu_t\right) &+ \hat{h}(\hat{x}_t) \geq \\
    \frac{\partial h}{\partial x}(f(\hat{x}_t)+g(\hat{x}_t)u_{t}) - \left\|\frac{\partial h}{\partial x} K_t c\right\|_2 \gamma & +\\
    \frac{1}{2}{tr}\left(\nu_{t}^{T}K_{t}^{T}\frac{\partial^{2}h}{\partial x^{2}}(\hat{x}_t)K_{t}\nu_{t}\right) &+\hat{h}(\hat{x}_t) \geq 0
    \end{aligned}
\end{equation*}
Hence, by Proposition 1, we have $Pr(x_{t} \in \mathcal{C}, 0\leq t\leq T  | \ ||x_{t}-\hat{x}_{t}||_{2} \leq \gamma) \geq (\frac{\hat{h}(x_0)}{\zeta})e^{-\zeta T}$. 
\end{proof}

Intuitively,  Eq. (\ref{eq:SCBF-2}) implies that as the state approaches the boundary, the control input is chosen such that the rate of increase of the barrier function decreases to zero. Hence Theorem \ref{theorem:SCBF} implies that if there exists an SCBF for a system, then the safety condition is satisfied with probability greater or equal to $(1-\epsilon)(\frac{\hat{h}(x_0)}{\zeta})e^{-\zeta T}$ when an EKF is used as an estimator and the control input is chosen at each time $t$ to satisfy (\ref{eq:SCBF-2}). We next present a probabilistic guarantee for HOCBFs within a finite time horizon.

\begin{Definition}[SCBF]
\label{Def:SCBF}
    The function $h$ is a \emph{Stochastic Control Barrier Function (SCBF)} of the system, if for all $\hat{x}_t$ satisfying $h(\hat{x}_t)>0$ there exists $u_t$ s.t. $\forall z_t$ with $\|z_t\|\leq \gamma$ \eqref{eq:SCBF-2} is satisfied. 
\end{Definition}

There may not exist a value of $u_t$ to satisfy \eqref{eq:SCBF-2} when $\frac{\partial h}{\partial x}g(x)=0$. To address this problem, CBF with high relative degree has been proposed in deterministic \cite{xiao2019control} and stochastic \cite{clark2021control} settings. 

Let $d=0,1, \ldots$ and define $d^{th}$ order differentiable
function $h^{d}(x)$ as
\begin{IEEEeqnarray*}{rCl}
    h^{0}(x) &=& h(x), \\
    h^{1}(x) &=& \frac{\partial h^{0}}{\partial x} \overline{f}(x,u)+\frac{1}{2} \operatorname{tr}\left(\sigma^{T}\left(\frac{\partial^{2} h^{0}}{\partial x^{2}}\right) \sigma\right)+h^{0}(x), \\
    &\vdots& \\
    h^{d+1}(x) &=& \frac{\partial h^{d}}{\partial x} \overline{f}(x,u)+\frac{1}{2} \operatorname{tr}\left(\sigma^{T}\left(\frac{\partial^{2} h^{d}}{\partial x^{2}}\right) \sigma\right)+h^{d}(x). 
\end{IEEEeqnarray*}
Define $\mathcal{C}^{d}=\left\{x: h^{d}(x) \geq 0\right\}$.
The following theorem provides sufficient conditions for safety of a high-degree system.
\begin{Theorem}
\label{th:hocbf}
Let $\overline{\mathcal{C}}=\bigcap_{d=0}^{d^{\prime}} \mathcal{C}^{d}$. Suppose that there exists $d^{\prime}$ such that $\frac{\partial h^{d^{\prime}}}{\partial x} g(x) u \neq 0$. If $u_{t}$ is chosen to satisfy 
\begin{equation}
\label{eq:HOCBF}
    \frac{\partial h^{d^{\prime}}}{\partial x}(\overline{f}(x,u))+\frac{1}{2} \operatorname{tr}\left(\sigma^{T} \frac{\partial^{2} h^{d^{\prime}}}{\partial x^{2}} \sigma\right) \geq-h^{d^{\prime}}(x) .
\end{equation}
Let $\zeta^{d}=\sup_{x\in \mathcal{C}^{d}}h^{d}(x)$. Then $Pr(x_{t} \in \mathcal{C}, 0\leq t\leq T) \geq \prod_{d=0}^{d^{\prime}}(\frac{h^{d}(x_0)}{\zeta^{d}})e^{-\zeta^{d} T}$ if $x_{0} \in \overline{\mathcal{C}}$.
\end{Theorem}
\begin{proof}
Suppose that $u_t$ satisfying the conditions of the theorem is chosen at each time $t$. Theorem 2 implies that $h^{d^{\prime}}\left(x_t\right) \geq 0$ when $0\leq t\leq T$ with probability greater or equal to $(\frac{h^{d^{\prime}}(x_0)}{\zeta^{d^{\prime}}})e^{-\zeta^{d^{\prime}} T}$. 
By definition of $h^{d}(x)$ we have that $\frac{\partial h^{d^{\prime}-1}}{\partial x} g(x) u = 0$. We also have $Pr(x_t\in \mathcal{C}^{d^{\prime}-1},0\leq t\leq T | x_t\in \mathcal{C}^{d^{\prime}}) \geq (\frac{h^{d^{\prime}-1}(x_0)}{\zeta^{d^{\prime}-1}})e^{-\zeta^{d^{\prime}-1} T}$. 
Proceeding inductively, we have $Pr(x_t\in \mathcal{C},0\leq t\leq T) \geq P$ where
\begin{multline*}
    P = \prod_{i=1}^{d^{\prime}}Pr(x_t\in \mathcal{C}^{i-1},0\leq t\leq T\ | x_t \in \mathcal{C}^{i})  \\
    \cdot Pr(x_t \in \mathcal{C}^{d^{\prime}}, 0\leq t\leq T). 
\end{multline*}
Finally, we have $Pr(x_t\in \mathcal{C},0\leq t\leq T)\geq \prod_{d=0}^{d^{\prime}} (\frac{h^{d}(x_0)}{\zeta^{d}})e^{-\zeta^{d} T}$.
\end{proof}
\begin{Definition}
    \label{Def:HOCBF}
    The function $h^{d^{\prime}}$ is a high-order CBF (HOCBF) of relative degree $d^{\prime}$ for system \eqref{eq:state-sde}-\eqref{eq:output-sde} if for all $x\in \overline{\mathcal{C}}$ there exists $u$ satisfying \eqref{eq:HOCBF}. 
\end{Definition}
The following theorem provides equivalent conditions for the existence of solutions to systems
of polynomial equations and inequalities. This allows us to verify sufficient and necessary conditions for safety.
\begin{Theorem}[Positivstellensatz \cite{parrilo2003semidefinite}]
\label{th:Positivstellensatz}
Let $\left(\phi_{j}\right)_{j=1, \ldots, s}$, $\left(\chi_{k}\right)_{k=1, \ldots, t}$, $\left(\psi_{\ell}\right)_{\ell=1, \ldots, u}$ be finite families of polynomials in $\mathbb{R}\left[x_{1}, \ldots, x_{n}\right]$. Denote by $C$ the cone generated by $\left(\phi_{j}\right)_{j=1, \ldots, s}$, $M$ the multiplicative monoid generated by $\left(\chi_{k}\right)_{k=1, \ldots, t}$, and $D$ the ideal generated by $\left(\psi_{\ell}\right)_{\ell=1, \ldots, u}$. Then, the following properties are equivalent:
\begin{enumerate}
    \item The set
    $$
    \left\{
    \begin{array}{l|ll}
    x \in \mathbb{R}^{n} &
    \begin{array}{ll}
    \phi_{j}(x) \geq 0, & j=1, \ldots, s \\
    \chi_{k}(x) \neq 0, & k=1, \ldots, t \\
    \psi_{\ell}(x)=0, & l=1, \ldots, u
    \end{array}
    \end{array}
    \right\}
    $$
    is empty.
    \item There exist $\phi \in C, \chi \in M, \psi \in D$ such that $\phi+\chi^{2}+\psi=0$.
\end{enumerate}
\end{Theorem}

\begin{Lemma}[Farkas' Lemma \cite{matousek2006understanding}]
\label{Lemma:Farkas}
Let $A$ be a real matrix with $m$ rows and $n$ columns, and let ${b} \in \mathbb{R}^n_{u}$ be a vector. One of the following conditions holds. 
i) The system of inequalities $A {x} \leq {b}$ has a solution. 
ii) There exists ${y}$ such that ${y} \geq \boldsymbol{0}$, ${y}^{T} A=\boldsymbol{0}^{T}$ and ${y}^{T} {b} < 0$, where $\boldsymbol{0}$ denote a zero vector.
\end{Lemma}



%% file: sections/SensorFault.tex
\section{Safe Control Under Sensor Faults and Attacks}
\label{sec:PM_sensorfault}
In this section, we consider the system under sensor faults and derive safety guarantees.

We consider a nonlinear control system whose output may be affected by one of $m$ sensor faults. The set of possible faults is indexed as $\{r_{1},\ldots,r_{m}\}$. Each fault $r_{i}$ maps to a set of affected observations $\mathcal{F}(r_{i}) \subseteq \{1,\ldots,q\}$. We assume that $\mathcal{F}(r_{i}) \cap \mathcal{F}(r_{j}) = \emptyset$ for $i \neq j$. Let $r \in \{r_{1},\ldots,r_{m}\}$ denote the index of the fault experienced by the system.
The system dynamics are described as 
\begin{eqnarray}
\label{eq:sfstate-sde}
dx_{t} &=& (f(x_{t})+g(x_{t})u_{t}) \ dt + \sigma_{t} \ dW_{t}\\
\label{eq:sfoutput-sde}
dy_{t} &=& (cx_{t} + a_{t}) \ dt + \nu_{t} \ dV_{t},
\end{eqnarray}
where the vector $a_{t} \in \mathbb{R}^{q}$ represents the impact of the fault and is constrained by $\mbox{supp}(a_{t}) \subseteq \mathcal{F}(r)$. Hence, if fault $r_{i}$ occurs, then the outputs of any of the sensors indexed in $\mathcal{F}(r_{i})$ can be arbitrarily modified by the fault. The sets $\mathcal{F}(r_{1}),\ldots,\mathcal{F}(r_{m})$ are known, but the value of $r_{i}$ is unknown.  In other words, the set of possible faults is known, but the exact fault that has occurred is unknown to the controller. 

To illustrate, we take an autonomous vehicle system as an example. Consider an autonomous system equipped with two Inertial Navigation System (INS) sensors, a Global Navigation Satellite System (GNSS) and one Light Detection And Ranging (LiDAR) system indexed as $\{1,2,3,4\}$ for localization measurements denoted as $y = \{y_{1},y_{2},y_{3},y_{4}\}$.
We consider three possible attacks: an attack on one of the INS sensors, an attack on another INS sensor, or a simultaneous GNSS/LiDAR spoofing attack. The corresponding fault patterns are given as $\mathcal{F}(r_{1})=\{1\}$, $\mathcal{F}(r_{2})=\{2\}$, $\mathcal{F}(r_{3})=\{3,4\}$. 

Define $\overline{c}_{i}$ to be the $c$ matrix with the corresponding rows indexed in $\mathcal{F}(r_{i})$ removed,
$\overline{y}_{t,i}$ to be equal to the vector $y_t$ with the entries indexed in $\mathcal{F}(r_{i})$ removed, and $\overline{\nu}_{t,i}$ to be the matrix $\nu_{t}$ with rows and columns indexed in $\mathcal{F}(r_{i})$ removed.
Define $\overline{c}_{i,j}$ to be the $c$ matrix with the corresponding rows indexed in $\mathcal{F}(r_{i})$ and $\mathcal{F}(r_{j})$ removed, where $i\neq j$. 

We make the following assumption for the sensor fault scenario. 
\begin{assumption}
\label{assumption:sf}
The system \eqref{eq:sfstate-sde}-\eqref{eq:sfoutput-sde} and the sensor fault patterns $\mathcal{F}(r_1),\ldots,\mathcal{F}(r_m)$ satisfy the conditions:
\begin{enumerate}
    \item The system is controllable. 
    \item For each $i,j \in \{1,\ldots,m\}$, the pair\\ $[\frac{\partial \overline{f}}{\partial x}(x,u), \overline{c}_{i,j}]$ is uniformly detectable.
\end{enumerate}
\end{assumption}
\noindent\textbf{Problem Statement:} Given a finite time $T$, a safe set $\mathcal{C}$ defined in \eqref{eq:safe-region} and a parameter $\epsilon \in (0,1)$, compute a control policy that, at each time $t$, maps the sequence $\{y_{t^{\prime}} : t^{\prime} \in [0,t)\}$ to an input $u_{t}$ such that, for any fault $r \in \{r_{1},\ldots,r_{m}\}$, $Pr(x_{t} \in \mathcal{C}, \ 0< t<T) \geq (1-\epsilon)\mathcal{T}(T)$, for some function $\mathcal{T}:\mathbb{R}^{n}\rightarrow(0,1)$. 

\subsection{Sensor FTC Strategy Definition}
We propose a CBF-based strategy with safety guarantees for a system satisfying Assumption \ref{assumption:sf}. The strategy that accommodates sensor faults and attacks is an FTC in a passive manner. The goal of FTC is to ensure the robustness of the control system to accommodate multiple component faults without striving for optimal performance for any specific fault condition.

The intuition behind our approach is as follows. Since we do not know the fault pattern $r$, we construct estimators excluding faulty sensors by maintaining $m$ EKFs. Each EKF corresponds to a different possible fault pattern in $\{r_1,\dots r_m\}$. We ensure safety with desired probability by defining $m$ corresponding SCBFs, each of which results in a different linear constraint on the control input. 

The potential drawback is that the safety guarantees of Theorem \ref{theorem:SCBF} rely on the existence of a control input satisfying the safety constraint at each time-step.
This assumption may not hold for two reasons. Firstly, feasible control input $u$ may not exist when $\frac{\partial h}{\partial x}g(x)=0$, since $u$ does not affect states $x$. 
Secondly, a feasible solution may not exist when faulty sensor measurements cause the state estimates to diverge. 
To address the first reason, we define high-order SCBFs such that for $d$-th degree $\frac{\partial h^d}{\partial x}g(x)\neq0$. Then, we choose control input $u$ to satisfy constraints constructed by high-order SCBFs.
To address the second reason, we define a set of $\binom{m}{2}$ EKFs to resolve conflicts between the constraints. Each EKF estimator omits all sensors affected by either fault $r_{i}$ or fault $r_{j}$ for some $i,j \in \{1,\ldots,m\}$, $i\neq j$. These estimators will be used to resolve any deviations between the state estimates from sensors $\{1,...,m\}\backslash\mathcal{F}(r_i)$ and $\{1,...,m\}\backslash\mathcal{F}(r_j)$.

Let $\mathcal{C}_{\gamma}:=\{x:\hat{h}^{d}(x)\geq 0\}$ where $\hat{h}^{d}(x)=h^{d}(x)-\bar{h}^{d}_{\gamma}$ and
\begin{equation}
\label{eq:bar_h_d_gamma}
\bar{h}^{d}_{\gamma}=\sup_{x,x^{d,0}} \left\{h^{d}(x):\|x-x^{d,0}\|_{2} \leq \gamma \text{ and } h^{d}(x^{d,0})=0\right\}
\end{equation}
Let $\overline{\mathcal{C}}_{\gamma}=\bigcap_{d=0}^{d^{\prime}} \mathcal{C}_{\gamma}^{d}$. To ensure safety as defined in \eqref{eq:safe-region}, we need to show that $Pr(x_{t} \in \mathcal{C}, \ 0\leq t\leq T) \geq \prod_{d=0}^{d^{\prime}}(\frac{\hat{h}^{d}(x_0)}{\zeta^{d}})e^{-\zeta^{d} T}$ if $x_0\in \bar{\mathcal{C}}_{\gamma}$ given $x_{0} \in \Bar{\mathcal{C}}_{\gamma}$ and $||x_{t}-\hat{x}_{t}||_{2} \leq \gamma, \ \forall t$. 

\begin{proposition}
\label{prop:HOCBF}
For a system \eqref{eq:sfstate-sde}-\eqref{eq:sfoutput-sde} with safety region defined by \eqref{eq:safe-region}, 
suppose there exists $d^{\prime}$, such that $\frac{\partial h^{d^{\prime}}}{\partial x}g(x) \neq 0$.
Suppose that $u_{t}$ is chosen to satisfy 
\begin{multline}
\label{eq:hoscbf}
    \frac{\partial h^{d^{\prime}}}{\partial x}
    \bar{f}(\hat{x}_t,u_t)
    -\left\|\frac{\partial h^{d^{\prime}}}{\partial x}(\hat{x}_{t})K_{t} c\right\|_{2}\gamma \\
    + \frac{1}{2} tr\left(\nu_{t}^{T} K_{t}^{T} \frac{\partial^{2}h^{d^{\prime}}}{\partial x}(\hat{x}_t) K_{t}\nu_{t}\right)\geq -\hat{h}^{d^{\prime}}(\hat{x}_t). 
\end{multline}
Then $Pr(x_{t} \in \mathcal{C}, \ 0\leq t\leq T | \ ||x_{t}-\hat{x}_{t}||_{2} \leq \gamma \ \forall t) \geq \prod_{d=0}^{d^{\prime}}(\frac{\hat{h}^{d}(x_0)}{\zeta^{d}})e^{-\zeta^{d} T}$ if $x_0\in \bar{\mathcal{C}}_{\gamma}$. 
\end{proposition}
\begin{proof}
Given $||x_{t}-\hat{x}_{t}||_{2} \leq \gamma \ \forall t$, we suppose that $u_t$ is chosen to satisfy \eqref{eq:hoscbf}.  
By Theorem \ref{theorem:SCBF} and \eqref{eq:hoscbf}, we have $h^{d^{\prime}}\left(x_t\right) \geq 0$ when $0\leq t\leq T$ with probability greater or equal to $(\frac{\hat{h}^{d^{\prime}}(x_0)}{\zeta^{d^{\prime}}})e^{-\zeta^{d^{\prime}} T}$. By definition of relative degree, we have $\frac{\partial h^{d}}{\partial x}g(x)u = 0$ for $d<d'$. By definition of $h^{d'}(x_t)$, we have $\frac{\partial h^{d}}{\partial x} \overline{f}(x,0)+\frac{1}{2} \operatorname{tr}\left(\sigma^{T}\left(\frac{\partial^{2} h^{d}}{\partial x^{2}}\right) \sigma\right)+h^{d}(x)\geq 0$, where $d=d^{\prime}-1$. This implies $h^{d'-1}(x_t)\geq 0$. Similar to the proof of Theorem \ref{th:hocbf}, by proceeding inductively, we then have $Pr(x_{t} \in \mathcal{C}, \ 0\leq t\leq T | \ ||x_{t}-\hat{x}_{t}||_{2} \leq \gamma \ \forall t) \geq \prod_{d=0}^{d^{\prime}}(\frac{\hat{h}^{d}(x_0)}{\zeta^{d}})e^{-\zeta^{d} T}$ if $x_0\in \bar{\mathcal{C}}_{\gamma}$. 
\end{proof}

The function $h^{d'}$ ensures safety of the system with relative degree $d'$ by Proposition \ref{prop:HOCBF}. Hence we define the function as follows. 
\begin{Definition}
    \label{Def:HOSCBF}
    The function $h^{d'}$ is a \emph{high-order SCBF (HOSCBF)} of relative degree $d'$ for system \eqref{eq:state-sde}-\eqref{eq:output-sde} if for all $\hat{x}_t\in \overline{\mathcal{C}}$ there exists $u_t$ satisfying \eqref{eq:hoscbf}. 
\end{Definition}

We next present a scheme to resolve conflicts between constraints in the case of faults and attacks. 
Let $\overline{R}_{t,i} = \overline{\nu}_{t,i}\overline{\nu}_{t,i}^{T}$ and $K_{t,i} = \overline{P}_{t,i}\overline{c}_{i}^{T}(\overline{R}_{t,i})^{-1}$. Here $\overline{P}_{t,i}$ is the solution to the Riccati differential equation $$\frac{d\overline{P}_{t,i}}{dt} = F_{t,i}\overline{P}_{t,i} + \overline{P}_{t,i}F_{t,i}^{T} + Q_{t}-\overline{P}_{t,i}\overline{c}_{i}^{T}\overline{R}_{t,i}^{-1}\overline{c}_{i}\overline{P}_{t,i}$$ with $F_{t,i} = \frac{\partial \overline{f}}{\partial x}(\hat{x}_{t,i},u_{t})$.  Define a set of $m$ EKFs with estimates denoted $\hat{x}_{t,i}$ via 
\begin{equation}
\label{eq:EKF-i}
d\hat{x}_{t,i} = (f(\hat{x}_{t,i}) + g(\hat{x}_{t,i})u_{t}) \ dt + K_{t,i}(d\overline{y}_{t,i}-\overline{c}_{i}\hat{x}_{t,i} \ dt).
\end{equation}
Each of these EKFs represents the estimate obtained by removing the sensors affected by fault $r_{i}$. Furthermore, define $\overline{y}_{t,i,j}$, $\overline{\nu}_{t,i,j}$, $\overline{c}_{i,j}$, $\overline{R}_{t,i,j}$, and $K_{t,i,j}$ in an analogous fashion with entries indexed in $\mathcal{F}(r_{i}) \cup \mathcal{F}(r_{j})$ removed. We assume throughout that the $\overline{R}$ matrices are invertible. We then define a set of ${m \choose 2}$ estimators $\hat{x}_{t,i,j}$ as 
\begin{multline}
\label{eq:EKF-ij}
d\hat{x}_{t,i,j} = (f(\hat{x}_{t,i,j}) + g(\hat{x}_{t,i,j})u_{t}) \ dt \\
+ K_{t,i,j}(d\overline{y}_{t,i,j}-\overline{c}_{i,j}\hat{x}_{t,i,j} \ dt).
\end{multline}
When $\mathcal{F}(r_{i}) \cup \mathcal{F}(r_{j}) = \{1,\ldots,q\}$, the open-loop estimator is used for $\hat{x}_{t,i,j}$. 

We then select parameters $\gamma_{1},\ldots,\gamma_{m} \in \mathbb{R}_{+}$, and $\{\theta_{ij} : i < j\} \subseteq \mathbb{R}_{+}$.
The set of feasible control inputs is defined at each time $t$ using the following steps:
\begin{enumerate}
\item Define $Z_{t} = \{1, \ldots, m\}$. Define a collection of sets $\Omega_{i}$, $i \in Z_{t}$, by  
\begin{multline}
    \Omega_{i} 
    \triangleq \left\{u: \frac{\partial h_i^{d^{\prime}}}{\partial x}
    \left(\bar{f}(\hat{x}_{t,i},u_t)\right)
    -\left\|\frac{\partial h_{i}^{d^{\prime}}}{\partial x}(\hat{x}_{t})K_{t} c\right\|_{2} \gamma_i \right.\\
    \label{eq:FTHOCBF-constraints}
    \left.+ \frac{1}{2} tr\left(\nu_{t}^{T} K_{t}^{T} \frac{\partial^{2}h_{i}^{d^{\prime}}}{\partial x}(\hat{x}_{t,i}) K_{t}\nu_{t}\right)\geq -\hat{h}_{i}^{d^{\prime}}(\hat{x}_{t,i}) \right\}
\end{multline}
Select  $u_{t}$ satisfying $u_{t} \in \bigcap_{i \in Z_{t}}{\Omega_{i}}$. If no such $u_{t}$ exists, there exists conflicts between constraints, i.e., $\exists i,j$, $i\neq j$ s.t. $\Omega_{i}\cap \Omega_{j}=\emptyset$. Then go to Step 2.
\item For each $i,j$ with $||\hat{x}_{t,i}-\hat{x}_{t,j}||_{2} > \theta_{ij}$, set $Z_{t} = Z_{t} \setminus \{i\}$ (resp. $Z_{t} = Z_{t} \setminus \{j\}$) if $||\hat{x}_{t,i}-\hat{x}_{t,i,j}||_{2} > \theta_{ij}/2$ (resp. $||\hat{x}_{t,j}-\hat{x}_{t,i,j}||_{2} > \theta_{ij}/2$).
If  $\bigcap_{i \in Z_{t}}{\Omega_{i}} \neq \emptyset$,  then select $u_{t} \in \bigcap_{i \in Z_{t}}{\Omega_{i}}$. Else, go to Step 3. This step resolves conflicts between estimations by comparing the difference between estimations against thresholds $\theta_{ij}$.
\item Remove the indices $i$ from $Z_{t}$ corresponding to the estimators with the largest residue values $\overline{y}_{t,i}-\overline{c}_{i}\hat{x}_{t,i}$ until there exists $u_{t} \in \bigcap_{i \in Z_{t}}{\Omega_{i}}$.
\end{enumerate}

We next provide sufficient conditions for this control policy to guarantee safety.
\begin{Theorem}
\label{theorem:FT-HOCBF}
Given $x_0\in \bar{\mathcal{C}}_{\gamma}$, define 
\begin{equation*}
\overline{h}^{d}_{\gamma_{i}} 
= \sup_{x,x^{0}}{\left\{h^{d}(x) : ||x-x^{d,0}||_{2} \leq \gamma_{i}\right.} \\
\left. \text{ and } h^{ d}(x^{d,0})=0\right\}
\end{equation*}
and $\hat{h}^{d}_{i}(x) = h^{d}(x)-\overline{h}^{d}_{\gamma_{i}}$. 
Suppose $\gamma_{1},\ldots,\gamma_{m}$, and $\theta_{ij}$ for $i < j$ are chosen such that the following conditions are satisfied: 
\begin{enumerate}
\item Define  $\Lambda^{d}_{i}(\hat{x}_{t,i}) = \frac{\partial h^{d}_{i}}{\partial x}(\hat{x}_{t,i})g(\hat{x}_{t,i})$.
For all $i,j \in Z_{t}$ with $||\hat{x}_{t,i}-\hat{x}_{t,j}||_{2} \leq \theta_{ij}$, there exists $u$ such that
\begin{equation}
\label{eq:FT-CBF2}
\Lambda^{d^{\prime}}_{i}(\hat{x}_{t,i})u > 0
\end{equation}
for all $i \in Z_{t}^{\prime}$.
\item For each $i$, when $r=r_{i}$, 
\begin{multline}
\label{eq:FT-CBF3}
Pr(||\hat{x}_{t,i}-\hat{x}_{t,i,j}||_{2} \leq \frac{\theta_{ij}}{2} \ \forall j,\\ ||\hat{x}_{t,i}-x_{t}||_{2} \leq \gamma_{i} \ \forall t)
 \geq 1-\epsilon.
\end{multline}
\end{enumerate}
Then $Pr(x_{t} \in \mathcal{C}, \ 0\leq t\leq T) \geq (1-\epsilon)\prod_{d=0}^{d^{\prime}}(\frac{\hat{h}^{d}(x_0)}{\zeta^{d}})e^{-\zeta^{d} T}$ for any fault pattern $r \in \{r_{1},\ldots,r_{m}\}$. 
\end{Theorem}
\begin{proof}
Suppose that the fault $f=f_i$. We will show that, if $||\hat{x}_{t,i}-x_{t}||_{2} \leq \gamma_{i}$ and $||\hat{x}_{t,i}-\hat{x}_{t,i,j}||_{2} \leq \theta_{ij}/2$ for all $t$, then  $u_{t} \in \Omega_{i}$ holds. Hence $x_{t} \in \mathcal{C}$ for $0\leq t\leq T$ with probability greater or equal to $\prod_{d=0}^{d^{\prime}}(\frac{\hat{h}^{d}(x_0)}{\zeta^{d}})e^{-\zeta^{d} T}$ by Proposition \ref{prop:HOCBF}. 

At time $t$, suppose that $\hat{h}^{d^{\prime}}_{i}(\hat{x}_{t,i}) \geq 0$, and that  $||\hat{x}_{t,i}-\hat{x}_{t,i,j}||_{2} \leq \theta_{ij}/2$. We consider three cases, namely (i) $||\hat{x}_{t,j} - \hat{x}_{t,k}||_{2} \leq \theta_{jk}$ for all $j,k \in Z_{t}$, (ii) $||\hat{x}_{t,i}-\hat{x}_{t,j}||_{2} \leq \theta_{ij}$ for all $j \in Z_{t}$, but there exist $j,k \in Z_{t} \setminus \{i\}$ such that $||\hat{x}_{t,j}-\hat{x}_{t,k}||_{2} > \theta_{jk}$, and (iii) $||\hat{x}_{t,i}-\hat{x}_{t,j}||_{2} > \theta_{ij}$ for some $j \in Z_{t}$.

\noindent \emph{\underline{Case (i):}} We will show that there exists $u \in \cap_{j \in Z_{t}}{\Omega_{j}}$, and hence in particular $u_{t}$ satisfies $\Omega_{i}$. Each $\Omega_{j}$ can be written in the form 
\begin{equation}
\label{eq:FT-CBF-case1}
\Omega_{j} = \{u : \Lambda^{d^{\prime}}_{j}(\hat{x}_{t,j})u_{t} \geq \overline{\omega}^{d^{\prime}}_{j}\}
\end{equation}
where  $\overline{\omega}^{d^{\prime}}_{j}$ is a real number that does not depend on $u_{t}$. Under the assumption 1) of the theorem, there exists $u$ satisfying (\ref{eq:FT-CBF2}) for all $i \in Z_{t}^{\prime}$. 
Choose $$u_{t} = \left(\max_{j}{\{|\overline{\omega}^{d^{\prime}}_{j}|}\}/||u||_{2}\right)u.$$ This choice of $u_{t}$ satisfies $u_{t} \in \bigcap_{j \in Z_{t}}{\Omega_{j}}$, in particular $u_{t} \in \Omega_{i}$.

\noindent \emph{\underline{Case (ii):}} In this case, Step 2 of the procedure is reached and constraints $\Omega_{j}$ are removed until all indices in $Z_{t}$ satisfy $||\hat{x}_{t,j}-\hat{x}_{t,k}||_{2} \leq \theta_{jk}$. Since $||\hat{x}_{t,i}-\hat{x}_{t,j}||_{2} \leq \theta_{ij}$ already holds for all $j \in Z_{t}$, $i$ will not be removed from $Z_{t}$ during this step. After Step 2 is complete, the analysis of Case (i) holds and there exists a $u$ which satisfies all the remaining constraints, including $\Omega_{i}$.

\noindent \emph{\underline{Case (iii):}} Suppose $j$ satisfies $||\hat{x}_{t,i}-\hat{x}_{t,j}||_{2} > \theta_{ij}$. We have 
\begin{eqnarray}
\nonumber
\theta_{ij} &<& ||\hat{x}_{t,i}-\hat{x}_{t,i,j} + \hat{x}_{t,i,j} - \hat{x}_{t,j}||_{2} \\
\label{eq:FT-CBF-case3-1}
&\leq& ||\hat{x}_{t,i}-\hat{x}_{t,i,j}||_{2} + ||\hat{x}_{t,i,j}-\hat{x}_{t,j}||_{2} \\
\label{eq:FT-CBF-case3-2}
&\leq& \theta_{ij}/2 + ||\hat{x}_{t,i,j}-\hat{x}_{t,j}||_{2}
\end{eqnarray}
where Eq. (\ref{eq:FT-CBF-case3-1}) follows from the triangle inequality and (\ref{eq:FT-CBF-case3-2}) follows from the assumption that $||\hat{x}_{t,i}-\hat{x}_{t,i,j}||_{2} \leq \theta_{ij}/2$. Hence $||\hat{x}_{t,j}-\hat{x}_{t,i,j}||_{2} > \theta_{ij}/2$ and $j$ is removed from $Z_{t}$. By applying this argument to all such indices $j$, we have that $i$ is not removed during Step 2 of the procedure, and thus the analyses of Cases (i) and (ii) imply that $u_{t} \in \Omega_{i}$.

From these cases, we have that $\Omega_{i}$ holds whenever $\hat{h}^{d}_{i}(\hat{x}_{t,i}) \geq 0$. Therefore, by Proposition \ref{prop:HOCBF}, we have
\begin{multline*}
Pr(x_{t} \in \mathcal{C}, \ 0\leq t\leq T | \  ||\hat{x}_{t,i}-\hat{x}_{t}||_{2} \leq \gamma_{i},\\ ||\hat{x}_{t,i}-\hat{x}_{t,i,j}||_{2} \leq \theta_{ij}/2 \ \forall t) \geq \prod_{d=0}^{d^{\prime}}(\frac{\hat{h}^{d}(x_0)}{\zeta^{d}})e^{-\zeta^{d} T}
\end{multline*}
 and $Pr(x_{t} \in \mathcal{C}, \ 0\leq t\leq T) \geq (1-\epsilon)\prod_{d=0}^{d^{\prime}}(\frac{\hat{h}^{d}(x_0)}{\zeta^{d}})e^{-\zeta^{d} T}$ by (\ref{eq:FT-CBF3}). 
\end{proof}

The bank of functions in Proposition \ref{theorem:FT-HOCBF} ensures the safety of the system with faulty components. Hence we define the functions as follows.
\begin{Definition}
    \label{Def:FTHOSCBF}
    The bank of functions $h^{d^{\prime}}_{1}, \ldots, h^{d^{\prime}}_{m}$ are \emph{Fault-Tolerant High Order Stochastic Control Barrier Functions (FT-HOSCBFs)} of relative degree $d^{\prime}$ for system \eqref{eq:state-sde}-\eqref{eq:output-sde} if conditions in Theorem \ref{theorem:FT-HOCBF} are satisfied. 
\end{Definition}

%% file: sections/FeasibilityVerification.tex
\subsection{Feasibility Verification}
\label{subsec:feasibility_verif}
In order for Theorem \ref{theorem:FT-HOCBF} to guarantee system safety, the linear constraint \eqref{eq:FT-CBF2} must hold for all time $t$. 
In what follows, we develop an SOS-based scheme to verify the feasibility of SCBF, FT-SCBF and FT-HOCBF constraints for both fault-free case and the case with sensor faults and attacks. 



We focus on verification for a constant-gain Kalman filter. In the case where the system is LTI with constant noise, the steady-state Kalman filter gain is optimal and hence satisfies the stochastic stability criteria by Theorem \ref{theorem:EKF}. In this subsection, we omit the time subscript of $x_t$, $\hat{x}_t$ and $z_t$, i.e., ($x$, $\hat{x}$ and $z$) to simplify the expression. We consider an LTI system described by \eqref{eq:state-sde} and \eqref{eq:output-sde}, where $f(x)=F$, $g(x)=G$, the matrices $R_t = R$ and $Q_t=Q$.
For an LTI system, $P_t$ is the covariance matrix for the estimation error and will converge to a steady-state value $P$. The Kalman filter has a constant gain given by $K = Pc^{T}R^{-1}$. 
We introduce an SOS-based approach to verify the feasibility for this case. 


\subsubsection{Verification for SCBF}
We first present the verification for an SCBF in an attack-free scenario, in which one SCBF-based safety constraint must be satisfied. We have the following initial result. 
\begin{proposition}
\label{prop:verify_scbf}
Suppose Assumption \ref{assumption:ekf} holds. The function $h(\hat{x})$ is a SCBF if and only if there is no $\hat{x}\in \mathcal{C}_{\gamma}$, $z\in\mathbb{R}^{n}$ satisfying
$\frac{\partial h}{\partial x}g(\hat{x}) = 0$ , $z^{T}z - \gamma^2 \leq 0$ and $\xi(\hat{x}) < 0$ where 
\begin{multline}
\label{eq:prop2cond}
    \xi(\hat{x}) =  \frac{\partial h}{\partial x}f(\hat{x}) +  \frac{1}{2}\mathbf{tr}\left(\nu^{T}K^{T}\frac{\partial^{2}h}{\partial x^{2}}(\hat{x})K\nu\right) - \\
\left\|\frac{\partial h}{\partial x}(\hat{x})K c \right\|_2 \gamma +  \hat{h}(\hat{x}).
\end{multline}
\end{proposition}
\begin{proof}
By Theorem \ref{theorem:SCBF}, the set $\mathcal{C}_{\gamma}$ is positive invariant given $\|x-\hat{x}\|\leq \gamma$ if for all time $t$ $u_t$ is chosen to satisfy \eqref{eq:SCBF-2} $\forall z_t$ with $\|z_t\|\leq \gamma$. By Definition \ref{Def:SCBF}, we have that $h(\hat{x})$ is a SCBF if and only if \eqref{eq:SCBF-2} holds for all $\hat{x}\in \mathcal{C}_{\gamma}:=\{x:\hat{h}(x)\geq 0\}$. If $\frac{\partial h}{\partial x}g(\hat{x})\neq 0$, we can choose $u$ s.t. 
\begin{multline*}
\frac{\partial h}{\partial x}g(\hat{x}) u \geq \sup_{\|z\|\leq \gamma}\left\{-\frac{\partial h}{\partial x}f(\hat{x}) - \right. \\ \left. \frac{1}{2}\mathbf{tr}\left(\nu^{T}K^{T}\frac{\partial^{2}h}{\partial x^{2}}(\hat{x})K\nu\right) + \left\|\frac{\partial h}{\partial x}(\hat{x})K c \right\|_2 \gamma -  \hat{h}(\hat{x}) \right\} .
\end{multline*}
Since $\|z\|\leq \gamma$ is a compact set, such a $u$ always exists. Hence, (4) fails if and only if $\exists\hat{x}$ and $z$ with $\|z\|\leq \gamma$ s.t. (i) $\frac{\partial h}{\partial x}g(\hat{x})=0$, and (ii) $\xi(\hat{x})<0$ hold simultaneously. 
\end{proof}

Based on the proposition, we can formulate the following conditions via the Positivstellensatz. 
\begin{Lemma}
A polynomial $h(\hat{x})$ is an SCBF for system (\ref{eq:state-sde})--(\ref{eq:output-sde}) 
if and only if there exist polynomials $\rho(\hat{x},z)$, sum-of-squares polynomials $q_S(\hat{x},z)$, integers $r_1$ such that
\begin{equation}
    \label{eq:verify_scbf}
    \phi(\hat{x},z) + \chi(\hat{x},z) + \psi(\hat{x}) = 0,
\end{equation}
and
\begin{align*}
    \phi(\hat{x},z) &= \sum_{S\subseteq \{1,\ldots,3\}} q_{S}(\hat{x},z)\prod_{i\in S}\phi_i(\hat{x},z)  \\
    \chi(\hat{x},z) &=  \left(\xi(\hat{x})\right)^{2 r_1}  \\
    \psi(\hat{x})  &= \sum_{i=1}^{m}\rho_i(\hat{x},z)\left[\frac{\partial h}{\partial x} g(\hat{x})\right]_{i},
\end{align*}
where $\phi_1(\cdot) = -\xi(\hat{x})$, $\phi_2(\cdot)= -z^{T}z+\gamma^2$ and $\phi_3(\cdot) =\hat{h}(\hat{x})$.
\end{Lemma}
\begin{proof}
By Proposition \ref{prop:verify_scbf}, we have $h(\hat{x})$ is an SCBF iff there exist no $\hat{x}$, $z$ such that$\frac{\partial h}{\partial x}g(\hat{x})=0$, $z^{T} z -\gamma^2 \leq 0$ and $-\xi(\hat{x})>0$. The latter two conditions are equivalent to $-z^{T} z +\gamma^2 \geq 0$, and $-\xi(\hat{x}) \geq 0$, $\xi(\hat{x}) \neq 0$. 
These conditions are equivalent to \eqref{eq:verify_scbf} by the Positivstellensatz.
\end{proof}


\subsubsection{Verification for FT-SCBF}
We now extend the result into the case where sensors may experience faults and attacks. Specifically, we consider a nonlinear control system whose output may be affected by one of $m$ sensor faults described by \eqref{eq:sfstate-sde} and \eqref{eq:sfoutput-sde}. 

In this case, we need to verify the feasibility of $u$ to satisfy $m$ SCBF constraints under $m$ possible sensor faults.
To achieve this, we extend Proposition \ref{prop:verify_scbf} to verify the feasibility of a set of SCBF constraints via Farkas' Lemma. 

\begin{corollary}
\label{coro:Farkas}
Define $A(x)$ and $\Xi(x)$ as follows.
\begin{equation}
    \begin{split}
    \label{coro1_AXi}
        A(x) &= \left( A_1(x) \ldots A_m(x) \right)^{T} \\
        \Xi(x) &= [\xi_1(x) \ldots \xi_m(x)]^T
    \end{split}
\end{equation}
Control input $u$ that is chosen to satisfy a set of linear constraints can be written as 
\begin{equation}
    \label{eq:coro1_u}
    A(x) u \leq \Xi(x,z). 
\end{equation}
By Farkas's Lemma, the system $A(x) u \leq \Xi(x)$ has a solution $u\in\mathbb{R}^{p}$, if and only if there does not exist $y\in\mathbb{R}^{m}$ such that
\begin{equation}
    \label{eq:coro1_y}
    A^{T}(\hat{x})y=0, \ y\geq 0, \ \Xi^{T}(\hat{x})y<0.
\end{equation}
\end{corollary}

We define $A(x)$ and $\Xi(x)$ as follows
\begin{equation*}
    \begin{split}
        A(\hat{x}) &= \left( -\frac{\partial h}{\partial x} g(\hat{x}_1) \ldots -\frac{\partial h}{\partial x} g(\hat{x}_{m}) \right)^{T}, \\
        \Xi(\hat{x}) &= [\xi(\hat{x}_1) \ldots \xi(\hat{x}_{m})]^T.
    \end{split}
\end{equation*}

\begin{proposition}
\label{prop:verify_ftscbf}
Suppose Assumption \ref{assumption:ekf} and conditions in Theorem \ref{theorem:FT-HOCBF} hold.
There exists a feasible solution $u$ satisfying a set of $m$ SCBF constraints if and only if there is no $\hat{x}_1, \ldots, \hat{x}_{m}\in \mathcal{C}_{\gamma}$, $z_1, \ldots, z_{m}\in \mathbb{R}^n$ and $y\in \mathbb{R}^{m}$ satisfying $z_j^{T} z_j -\gamma^2 \leq 0$, $(\hat{x}_j-\hat{x}_k)^T (\hat{x}_j-\hat{x}_k)- \gamma^2 \leq 0$, $A^{T}(\hat{x}_j)y=0$, $y\geq 0$ and $\Xi^{T}(\hat{x}_j)y<0$, $\forall j,k\in\{1,\ldots,m\}$.
\end{proposition}
\begin{proof}
By Definition \ref{Def:SCBF}, we have $\hat{h}(\hat{x}_j) = h(\hat{x}_j)-\overline{h}_{\gamma}\geq 0$ for all $\hat{x}_j\in \mathcal{C}_{\gamma}$. By Theorem \ref{theorem:EKF}, we have $z_j^{T} z_j -\gamma^2 \leq 0$ for all $j$. 
By Theorem \ref{theorem:FT-HOCBF}, we have $(\hat{x}_j-\hat{x}_k)^T (\hat{x}_j-\hat{x}_k)\leq \gamma^2$, $\forall j,k\in\{1,\ldots,m\}$. 
For the case where $\hat{h}(\hat{x})=0$, $u$ can be chosen to satisfy \eqref{eq:SCBF-2} for all $j$. By Corollary \ref{coro:Farkas}, we have the existence of $u$ if and only if there does not exist $y\in\mathbb{R}^{m}$ such that \eqref{eq:coro1_y} hold.
Conversely, if for some $\hat{x}_0$, $\hat{x}_1$, $z_0$ and $y_0$ satisfying $\frac{\partial h}{\partial x}g(\hat{x}_0) = 0$, 
$(\hat{x}_0-\hat{x}_1)^T (\hat{x}_0-\hat{x}_1) - \gamma^2 \leq 0$, $z_0^{T} z_0 -\gamma^2 \leq 0 $, $A^{T}(\hat{x}_0)y_0=0$, $y_0\geq 0$ and $\Xi^{T}(\hat{x}_0)y_0<0$, the set $\mathcal{C}$ is not positive invariant.
\end{proof}

Then, we can formulate the following conditions via the Positivstellensatz.

\begin{Lemma}
\label{Lemma:verify_ftscbf}
There exists a feasible solution $u$ satisfying a set of $m$ SCBF constraints if and only if there exist polynomials $\rho(\hat{x}_j,y,z_j)$, sum-of-squares polynomials $q(\hat{x}_j,y,z_j)$, integers $s=4m+m^2$, $r_1,\ldots,r_m$ such that
\begin{equation}
    \label{eq:verify_ftscbf}
    \phi(\hat{x}_j,\hat{x}_k,y,z_j) + \chi(\hat{x}_j,\hat{x}_k,y,z_j) + \psi(\hat{x}_j,\hat{x}_k,y,z_j) = 0, 
\end{equation}
and
\begin{align*}
    \phi(\cdot) =& \sum_{S\subseteq \{1,\ldots,s\}} q_{S}(\hat{x}_j,\hat{x}_k,y,z_j)\prod_{i\in S}\phi_i(\hat{x}_j,\hat{x}_k,y,z_j)  \\
    \chi(\cdot) =& \prod_{\forall j\in\{1,\ldots,m\}}\left(-\Xi^{T}(\hat{x}_j)y\right)^{2 r_j}\\
    \psi(\cdot)  =& \sum_{j=1}^{m}\left( \sum_{i=1}^{p}\rho^0_i(\hat{x}_j,\hat{x}_k,y,z_j)\left[A^T(\hat{x}_j) y\right]_{i} \right),
\end{align*} 
where $\phi_{\{1,\ldots,m\}}(\cdot)=-\Xi^{T}(\hat{x}_j)y_j$,
$\phi_{\{m+1,\ldots,2m\}}(\cdot)=\hat{h}(\hat{x}_j)$,
$\phi_{\{2m+1,\ldots,3m\}}(\cdot)=-z_j^{T}z_j+\gamma^2$, $\phi_{\{3m+1,\ldots,4m\}}(\cdot)=y_j$ and $\phi_{\{4m+1\ldots4m+m^2\}}(\cdot)=-(\hat{x}_j-\hat{x}_k)^T (\hat{x}_j-\hat{x}_k)+ \gamma^2$. 
\end{Lemma}
\begin{proof}
By Proposition \ref{prop:verify_ftscbf}, we have $h(\hat{x})$ is an SCBF if and only if there exist no $\hat{x}_1, \ldots, \hat{x}_{m}$, $z_1, \ldots, z_{m}$ and $y$ satisfying $(\hat{x}_j-\hat{x}_k)^T (\hat{x}_j-\hat{x}_k)- \gamma^2 \leq 0$, $\forall j,k\in\{1,\ldots,m\}$, $z_j^{T} z_j -\gamma^2 \leq 0 \ \forall j$, $A^{T}(\hat{x})y_j=\boldsymbol{0}$, $y_j\geq 0$ and $\Xi^{T}(\hat{x})y_j<0$. 
The conditions are equivalent to $\forall j,k\in\{1,\ldots,m\}$,
\begin{equation*}
    \begin{split}
    &-z_j^{T} z_j +\gamma^2 \geq 0 , \\
    &-(\hat{x}_j-\hat{x}_k)^T (\hat{x}_j-\hat{x}_k)+ \gamma^2 \geq 0, \\ 
    &A^{T}(\hat{x}_j)y=\boldsymbol{0}, \ y\geq 0, \ -\Xi^{T}(\hat{x}_j)y\geq 0, \Xi^{T}(\hat{x}_j)y\neq 0
    \end{split}
\end{equation*}
These conditions are equivalent to \eqref{eq:verify_ftscbf} by the Positivstellensatz.
\end{proof}

\subsubsection{Verification for FT-SCBF with high relative degree}
We further extend the proposition \ref{prop:verify_ftscbf} and Lemma \ref{Lemma:verify_ftscbf} to verify the feasibility of a set of HOSCBF constraints. 

We define $A(x)$ and $\Xi(x)$ as follows
\begin{equation*}
    \begin{split}
        A(\hat{x}) =& \left( -\frac{\partial h^{d^{\prime}}}{\partial x} g(\hat{x}_1), \ldots, -\frac{\partial h^{d^{\prime}}}{\partial x} g(\hat{x}_{m}) \right)^{T}, \\
        \Xi(\hat{x}) =& [\xi_{1}(\hat{x}_1), \ldots, \xi_{m}(\hat{x}_{m})]^T, where \\
        \xi_{i}(\hat{x}) =&  \frac{\partial h^{d^{\prime}}}{\partial x}f(\hat{x}) +  \frac{1}{2}\mathbf{tr}\left(\nu^{T}K^{T}\frac{\partial^{2}h_{i}^{d^{\prime}}}{\partial x^{2}}(\hat{x})K\nu\right) - \\
        & \ \left\|\frac{\partial h_{i}^{d^{\prime}}}{\partial x}(\hat{x})K c\right\|_{2} \gamma_{i} +  \hat{h}_{i}^{d^{\prime}}(\hat{x}).
    \end{split}
\end{equation*}

\begin{proposition}
\label{prop:verify_fthocbf}
Suppose Assumption \ref{assumption:ekf} and conditions in Theorem \ref{theorem:FT-HOCBF} hold.
There exists a feasible solution $u$ satisfying a set of $m$ HOSCBF constraints with relative degree $d$ if and only if there is no $\hat{x}_1, \ldots, \hat{x}_{m}\in \mathcal{C}_{\gamma}$, $z_1, \ldots, z_{m}\in \mathbb{R}^{n}$ and $y\in \mathbb{R}^{m}$ satisfying $\hat{h}^{d}(\hat{x_j})\geq 0, \ \forall j, \forall d\leq d^{\prime} $, $(\hat{x}_j-\hat{x}_k)^T (\hat{x}_j-\hat{x}_k)-  \gamma^2 \leq 0$, $\forall j,k\in\{1,\ldots,m\}$, $z_j^{T} z_j -\gamma^2 \leq 0 \ \forall j$, $A^{T}(\hat{x})y=0$, $y\geq 0$ and $\Xi^{T}(\hat{x})y<0$. 
\end{proposition}
\begin{proof}
By Theorem \ref{theorem:EKF}, we have $z_j^{T} z_j -\gamma^2 \leq 0$ for all $j$ with $m$ EKFs. 
By the Definition \ref{Def:HOSCBF}, we have $\hat{h}^{d}(\hat{x}_j) = h^{d}(\hat{x}_j)-\overline{h}_{\gamma}\geq 0$ for all $\hat{x}_j\in \overline{\mathcal{C}}$, if and only if the following three conditions are satisfied. For all $\hat{x}_j\in \overline{\mathcal{C}}$, $\hat{h}^{d}(\hat{x})\geq0$ for all $d\leq d^{\prime}$. Next, by Theorem \ref{theorem:FT-HOCBF}, $(\hat{x}_j-\hat{x}_k)^T (\hat{x}_j-\hat{x}_k)\leq \gamma^2$, $\forall j,k\in\{1,\ldots,m\}$. Moreover, $\frac{\partial h^{d^{\prime}}}{\partial x}g(\hat{x}_j)\neq 0$ and $u$ are chosen to satisfy \eqref{eq:hoscbf} for all $j$. By Corollary \ref{coro:Farkas}, we have a solution $u\in\mathbb{R}^{p}$ exists, if and only if there does not exist $y\in\mathbb{R}^{m}$ such that \eqref{eq:coro1_y} holds. 
Conversely, if for some $\hat{x}_0$, $\hat{x}_1$, $z_0$ and $y_0$ satisfying $\hat{h}^{d}(\hat{x}_0) \geq 0$, $(\hat{x}_0-\hat{x}_1)^T (\hat{x}_0-\hat{x}_1)- \gamma^2 \leq 0$, $z_0^{T} z_0 -\gamma^2 \leq 0 $, $A^{T}(\hat{x})y_0=0$, $y_0\geq 0$ and $\Xi^{T}(\hat{x})y_0<0$, the set $\mathcal{C}$ is not positive invariant. 
\end{proof}

\begin{Lemma}
\label{Lemma:verify_fthocbf}
There exists a feasible solution $u$ satisfying a set of $m$ HOSCBF constraints 
if and only if there exist polynomials $\rho(\hat{x}_j,\hat{x}_k,y,z_j)$, sum-of-squares polynomials $q_S(\hat{x}_j,\hat{x}_k,y,z_j)$, integers $d^{\prime}_1,\ldots,d^{\prime}_m$, $s=3m+m^2+\sum_{j=1}^{m}d^{\prime}_j$, $r_1,\ldots,r_m$ such that
\begin{equation}
    \label{eq:verify_fthocbf}
    \phi(\hat{x}_j,\hat{x}_k,y,z_j) + \chi(\hat{x}_j,\hat{x}_k,y,z_j) + \psi(\hat{x}_j,\hat{x}_k,y,z_j) = 0, 
\end{equation}
and
\begin{align*}
    \phi(\cdot) =& \sum_{S\subseteq \{1,\ldots,s\}} q_{S}(\hat{x}_j,\hat{x}_k,y,z_j)\prod_{i\in S}\phi_i(\hat{x}_j,\hat{x}_k,y,z_j)  \\
    \chi(\cdot) =& \prod_{\forall j\in\{1,\ldots,m\}}\left(-\Xi^{T}(\hat{x}_j)y\right)^{2 r_j}\\
    \psi(\cdot)  =& \sum_{j=1}^{m}\left(\sum_{i=1}^{p}\rho^0_i(\hat{x}_j,\hat{x}_k,y,z_j)\left[A^T(\hat{x}_j) y\right]_{i} \right)
\end{align*}
where $\phi_{\{1,\ldots,m\}}(\cdot)=-\Xi^{T}(\hat{x}_j)y$, $\phi_{\{m+1,\ldots,2m\}}(\cdot)=y_j$, $\phi_{\{2m+1,\ldots,3m\}}(\cdot)=-z_j^{T}z_j+\gamma^2$,$\phi_{\{3m+1\ldots3m+m^2\}}(\cdot)=-(\hat{x}_j-\hat{x}_k)^T (\hat{x}_j-\hat{x}_k)+ \gamma^2$ and for $d_j\in\{0,\ldots,d^{\prime}_j\}$, $\phi_{\{3m+m^2+1,\ldots,3m+m^2+\sum_{j=1}^{m}d^{\prime}_j\}}(\cdot)=\hat{h}^{d_j}(\hat{x}_j)$.
\end{Lemma}
\begin{proof}
By proposition \ref{prop:verify_fthocbf}, we have $h^{d}(\hat{x})$ are HOSCBFs if and only if there exist no $\hat{x}_1, \ldots, \hat{x}_{m}\in \mathcal{C}_{\gamma}$, $z_1, \ldots, z_{m}\in \mathbb{R}^{n}$ and $y\in \mathbb{R}^{m}$ satisfying $\hat{h}^{d}(\hat{x}_j)\geq0, \ \forall j, \forall d\leq d^{\prime}$, $(\hat{x}_j-\hat{x}_k)^T (\hat{x}_j-\hat{x}_k)- \gamma^2 \leq 0$, $\forall j,k\in\{1,\ldots,m\}$, $z_j^{T} z_j -\gamma^2 \leq 0 \ \forall j$, $A^{T}(\hat{x})y=0$, $y\geq 0$ and $\Xi^{T}(\hat{x})y<0$. 
The conditions are equivalent to $\forall j,k\in\{1,\ldots,m\}$,
\begin{equation*}
    \begin{split}
    &\hat{h}^{d_j}(\hat{x}_j)\geq0, \forall d_j\leq d^{\prime}_j\\
    &-(\hat{x}_j-\hat{x}_k)^T (\hat{x}_j-\hat{x}_k)+ \gamma^2 \geq 0,  -z_j^{T} z_j +\gamma^2 \geq 0, \\
    &A^{T}(\hat{x}_j)y=\boldsymbol{0}, \ y\geq 0, \ -\Xi^{T}(\hat{x}_j)y\geq 0, \Xi^{T}(\hat{x}_j)y\neq 0
    \end{split}
\end{equation*}
These conditions are equivalent to \eqref{eq:verify_fthocbf} by the Positivstellensatz.
\end{proof}

%% file: sections/SF_joint.tex
\section{Joint Safety and Stability Under Sensor Faults and Attacks}
\label{sec:joint-safety-reach}
We next present a framework to ensure joint safety and stability for systems with sensor faults and attacks via CLFs and HOSCBFs. Such an approach has been widely used in fault-free scenarios. 

Define the goal set $\mathcal{G}\subseteq \mathcal{C}$ by $\mathcal{G} = \{x : w(x) \geq 0\}$ for some function $w$ for some equilibrium point $x_e\in G$, $f(x_e)=0$ and $g(x_e)=0$. Define  $\mathcal{\tau}(\mathcal{G})$ as the first time when $x_{t}$ reaches $\mathcal{G}$.

\textbf{Problem Statement: } Given a goal set $\mathcal{G}$, a safe set $\mathcal{C}$ and a parameter $\epsilon \in (0,1)$, compute a control policy that, at each time $t$, maps the sequence $\{y_{t^{\prime}} : t^{\prime} \in [0,t)\}$ to an input $u_{t}$ such that, given a finite stopping time $T$, for any fault $r \in \{r_{1},\ldots,r_{m}\}$, $Pr(x_{t} \in \mathcal{C}, \ 0\leq t\leq T) \geq (1-\epsilon)\mathcal{T}(T)$, for some function $\mathcal{T}:\mathbb{R}^{n}\rightarrow(0,1)$ and $Pr(\tau(\mathcal{G}) < \infty) > 1-\epsilon$.

\subsection{HOSCBF-CLF}
Our approach towards through asymptotically convergence to goal set $\mathcal{G}$ is through the use of \emph{stochastic Control Lyapunov Functions}. A function $V: \mathbb{R}^{n} \rightarrow \mathbb{R}_{\geq 0}$ is a stochastic CLF for the SDE (\ref{eq:state-sde}) if, for each $x_t$, we have
\begin{equation}
    \label{eq:CLF-def}
    \inf_{u}{\left\{\frac{\partial V}{\partial x}\overline{f}(x_t,u_t)+ \frac{1}{2}\mathbf{tr}\left(\sigma^{T}\frac{\partial^{2}V}{\partial x^{2}}\sigma\right)\right\}} < -\rho V(x_{t})^{\eta}
\end{equation}
for some $\rho > 0$ and $0<\eta <1$. 

The following result describes the stochastic stability of systems using CLFs with \cite[Theorem 3.1]{yin2011finite} providing sufficient conditions for the following result. As a preliminary, define $\tau(z) = \inf{\{t : V(x_{t}) \leq z\}}$.
\begin{proposition}[{\cite[Theorem 3.1]{yin2011finite}}]
\label{prop:CLF-stability}
Suppose there exists a $\overline{V}$ such that, whenever $V(x_{t}) \geq \overline{V}$, we choose $u_t$ to satisfy $$\frac{\partial V}{\partial x}f(x_{t}) + \frac{\partial V}{\partial x}g(x_{t})u_{t} + \frac{1}{2}\mathbf{tr}\left(\sigma^{T}\frac{\partial^{2}V}{\partial x^{2}}\sigma\right) < -\rho V(x_{t})^{\eta}$$ for some $\rho > 0$ and $0<\eta <1$. For $x\in \mathbb{R}^{n}\backslash \{x| V(x)=0\}$, $Pr(\tau(\overline{V}) < \infty | x_{0}=x) = 1$.
\end{proposition}

In the case with sensor faults and attacks, we consider a system with dynamics (\ref{eq:state-sde}) and an Extended Kalman Filter estimator $\hat{x}_{t}$. The following result is an extension of Proposition \ref{prop:CLF-stability} to this case.
\begin{Lemma}
    \label{lemma:CLF-stability-incomplete}
    Suppose that there exist constants $M > 0$ and $k \in \mathbb{N}$ such that, for any $x$ and $x^{\prime}$, $|V(x)-V(x^{\prime})| \leq M||x-x^{\prime}||_{2}^{k}$. Suppose $Pr(||\hat{x}_{t}-x_{t}||_{2} \leq \gamma \ \forall t) > 1-\epsilon$ and, at each time $t$ when $V(\hat{x}_{t}) > \overline{V}$, we have 
\begin{multline}
    \label{eq:stability-condition}	
    \frac{\partial V}{\partial x}(\hat{x}_{t})(f(\hat{x}_{t})+g(\hat{x}_{t})u_{t}) + \gamma ||\frac{\partial V}{\partial x}(\hat{x}_{t})K_{t}c||_{2} \\
    + \frac{1}{2}\mathbf{tr}\left(\nu_{t}^{T}K_{t}^{T}\frac{\partial^{2}V}{\partial x^{2}}(\hat{x}_{t})K_{t}\nu_{t}\right) < -\rho (V(\hat{x}_{t})+M \gamma^{k})^{\eta}.	
\end{multline}
Then $$Pr(\tau(\overline{V} + M\gamma^{k}) < \infty) > 1-\epsilon$$ for all $x_t \in \mathbb{R}^{n}\backslash \{x_t| V(x_t)=0\}$.
\end{Lemma}

\begin{proof}
	The dynamics of $\hat{x}_{t}$ are given by $$d\hat{x}_{t} = \overline{f}(\hat{x}_{t},u_{t}) + K_{t}c(x_{t}-\hat{x}_{t}) + K_{t} \nu_{t} dV_{t}.$$ If $||\hat{x}_{t}-x_{t}||_{2} \leq \gamma$, the differential generator $LV(\hat{x}_{t})$ satisfies 
	\begin{eqnarray*}
	LV(\hat{x}_{t}) &=& \frac{\partial V}{\partial x}(\hat{x}_{t})\overline{f}(\hat{x}_{t},u_{t}) + \frac{\partial V}{\partial x}(\hat{x}_{t})K_{t}c(x_{t}-\hat{x}_{t}) \\
		&& + \frac{1}{2}\mathbf{tr}\left(\nu_{t}^{T}K_{t}^{T}\frac{\partial^{2}V}{\partial x^{2}}(\hat{x}_{t})K_{t}\nu_{t}\right) \\
		&\leq& \frac{\partial V}{\partial x}(\hat{x}_{t})\overline{f}(\hat{x}_{t},u_{t}) +	\gamma||\frac{\partial V}{\partial x}(\hat{x}_{t})K_{t}c||_{2} \\
		&& + \frac{1}{2}\mathbf{tr}\left(\nu_{t}^{T}K_{t}^{T}\frac{\partial^{2}V}{\partial x^{2}}(\hat{x}_{t})K_{t}\nu_{t}\right).
		\end{eqnarray*}
		since 
		\begin{eqnarray*}
		\frac{\partial V}{\partial x}(\hat{x}_{t})K_{t}c(x_{t}-\hat{x}_{t}) &\leq& ||\frac{\partial V}{\partial x}(\hat{x}_{t})K_{t}c||_{2}||x_{t}-\hat{x}_{t}||_{2} \\
		&\leq& \gamma ||\frac{\partial V}{\partial x}(\hat{x}_{t})K_{t}c||_{2}.
		\end{eqnarray*}
    Since $|V(x_{t})-V(\hat{x}_{t})| \leq M||x_{t}-\hat{x}_{t}||_{2}^{k}$, we have $V(x_{t}) \leq V(\hat{x}_{t}) + M||x_{t}-\hat{x}_{t}||_{2}^{k}$ and $\rho(V(x_t))^{\eta}\leq \rho(V(\hat{x}_{t})+M\gamma ^{k})^{\eta}$, for some $\rho>0$ and $0<\eta <1$. 
  
    Hence, if (\ref{eq:stability-condition}) holds, then
    $$Pr\left(\inf{\{t : V(\hat{x}_{t}) \leq \overline{V}\}} < \infty \ | \ ||x_{t}-\hat{x}_{t}||_{2} \leq \gamma \ \forall t\right) = 1$$
    by Proposition \ref{prop:CLF-stability}.
    Since $|V(x_{t})-V(\hat{x}_{t})| \leq M||x_{t}-\hat{x}_{t}||_{2}^{k}$, we have $V(x_{t}) \leq V(\hat{x}_{t}) + M||x_{t}-\hat{x}_{t}||_{2}^{k}$, and so 
    \begin{eqnarray*}
		&Pr(V(x_{t}) > \overline{V} + M\gamma^{k} | \ ||x_{t}-\hat{x}_{t}||_{2} \leq \gamma \ \forall t)  \\
		&\leq Pr(V(\hat{x}_{t}) + M\gamma^{k}>\overline{V}+ M\gamma^{k}| \ ||x_{t}-\hat{x}_{t}||_{2} \leq \gamma \ \forall t) \\
		&= Pr(V(\hat{x}_{t}) > \overline{V} | \ ||x_{t}-\hat{x}_{t}||_{2} \leq \gamma \ \forall t)
    \end{eqnarray*}
    Hence $Pr(\tau(\overline{V}+M\gamma^{k}) < \infty | \ ||x_{t}-\hat{x}_{t}||_{2} \leq \gamma \ \forall t) = 1$ and thus $Pr(\tau(\overline{V} + M\gamma^{k}) < \infty) > 1-\epsilon$.
\end{proof}
Motivated by this result, we next state a control policy that combines CLFs and HOSCBFs to ensure safety and stability. At each time $t$, the set of feasible control actions is defined as follows:
\begin{enumerate}
\item Define $Y_{t}(\overline{V}) = \{j : V(\hat{x}_{t,j}) > \overline{V})$, and initialize $U_{t} = Y_{t}(\overline{V})$. Define a collection of sets $\Upsilon_{i}$, $i \in U_{t}$, by 
\begin{multline}
\label{eq:CLF-constraint}
\Upsilon_{i} \triangleq \left\{u : \frac{\partial V_{i}}{\partial x}\overline{f}(\hat{x}_{t,i},u) + \gamma_{i}||\frac{\partial V_{i}}{\partial x}(\hat{x}_{t,i})c||_{2} \right. \\
\left. + \frac{1}{2}\mathbf{tr}\left(\overline{\nu}_{t,i}^{T}K_{t,i}^{T}\frac{\partial^{2}V}{\partial x^{2}}(\hat{x}_{t,i})K_{t,i}\overline{\nu}_{t,i}\right) \right. \\
\left. < -\rho_i (V(\hat{x}_{t})+M \gamma^{k})^{\eta_i}\right\}
\end{multline}
for some $\rho_i$, $\eta_i$, $i=1,\ldots,m$. Select any $$u_{t} \in \left(\bigcap_{i \in Z_{t}}{\Omega_{i}}\right) \cap \left(\bigcap_{j \in U_{t}}{\Upsilon_{j}}\right),$$ where $\Omega_{i}$ is defined as in (\ref{eq:FTHOCBF-constraints}). If no such $u_{t}$ exists, go to Step 2.
\item For each $i,j$ with $||\hat{x}_{t,i}-\hat{x}_{t,j}||_{2} > \overline{\theta}_{ij}$, set $Z_{t} = Z_{t} \setminus \{i\}$ and $U_{t} = U_{t} \setminus \{i\}$ (resp. $Z_{t} = Z_{t} \setminus \{j\}$ and $U_{t} = U_{t} \setminus \{j\}$)  if $||\hat{x}_{t,i}-\hat{x}_{t,i,j}||_{2} > \theta_{ij}/2$ (resp. $||\hat{x}_{t,j}-\hat{x}_{t,i,j}||_{2} > \theta_{ij}/2$). If $$\left(\bigcap_{i \in Z_{t}}{\Omega_{i}}\right)\cap \left(\bigcap_{j \in U_{t}}{\Upsilon_{j}}\right) \neq \emptyset,$$ then select $u_{t}$ from this set. Else go to Step 3.
\item Remove the sets $\Omega_{i}$ and $\Upsilon_{i}$ corresponding to the estimators with the largest residue values until there exists a feasible $u_{t}$.
\end{enumerate}
This policy is similar to the HOSCBF-based approach of Section \ref{sec:PM_sensorfault}, with additional constraints to satisfy the stability condition. This leads to another $m$ linear inequalities. The following result gives sufficient conditions for safety and stability.

\begin{Theorem}
\label{theorem:CBF-CLF-safety}
Suppose that $\hat{h}^{d^{\prime}}_{1},\ldots,\hat{h}^{d^{\prime}}_{m}$,  $\gamma_{1},\ldots,\gamma_{m}$, $V$, $\overline{V}$, and $\theta_{ij}$ satisfy the constraints of Theorem \ref{theorem:FT-HOCBF}, as well as the following: (i) The function $V$ satisfies $\{x: V(x) \leq \overline{V} + M\gamma_{i}^{k}\} \subseteq \mathcal{G}$ for all $i$. (ii)   Define $\Gamma_{i}(\hat{x}_{t,i}) = \frac{\partial V_{i}}{\partial x}g(\hat{x}_{t,i})$. Let $X_{t}^{\prime} \subseteq X_{t}^{d^{\prime}}(\delta)$ and $Y_{t}^{\prime} \subseteq Y_{t}(\overline{V})$ be sets satisfying $||\hat{x}_{t,i}-\hat{x}_{t,j}||_{2} \leq \theta_{ij}$ for all $i \in X_{t}^{\prime}$ and $j \in Y_{t}^{\prime}$. Then there exists $u$ with 
\begin{equation}
\label{eq:CBF-CLF}	
\Lambda_{i}^{d^{\prime}}(\hat{x}_{t,i})u > 0, \quad \Gamma_{j}(\hat{x}_{t,j})u < 0
\end{equation}
for all $i \in X_{t}^{\prime}$ and $j \in Y_{t}^{\prime}$. If conditions (i) and (ii) hold, then $Pr(x_{t} \in \mathcal{C}, \ 0\leq t\leq T) \geq (1-\epsilon)\prod_{d=0}^{d^{\prime}}(\frac{\hat{h}^{d}(x_0)}{\zeta^{d}})e^{-\zeta^{d} T}$ and $Pr(\tau(\mathcal{G}) < \infty) > 1-\epsilon$ for any fault pattern $r \in \{r_{1},\ldots,r_{m}\}$, where $\mathcal{\tau}(\mathcal{G})$ is the first time when $x_{t}$ reaches $\mathcal{G}$. 
\end{Theorem}

\begin{proof}
Suppose there exists relative degree $d^{\prime}$. By the argument of Theorem \ref{theorem:FT-HOCBF}, $i \in X_{t}^{d^{\prime}}(\delta)$ implies that $\Omega_{i}$ is a constraint on $u_{t}$ at time $t$. An analogous argument yields that $\Upsilon_{i}$ is a constraint as well.
By selecting $u_{t}$ satisfying (\ref{eq:CBF-CLF}) at each time $t$, we have that $Pr(x_{t} \in \mathcal{C}, \ 0\leq t\leq T) \geq (1-\epsilon)\prod_{d=0}^{d^{\prime}}(\frac{\hat{h}^{d}(x_0)}{\zeta^{d}})e^{-\zeta^{d} T}$ by Theorem \ref{theorem:FT-HOCBF} and $Pr(\tau(\overline{V}+M\gamma^{k}) < \infty) > 1-\epsilon$ by Lemma \ref{lemma:CLF-stability-incomplete}.
Hence $Pr(\tau(\mathcal{G})< \infty) > 1-\epsilon$ by assumption (i) of the theorem. 	
\end{proof}

A controller that reaches a goal set defined by a function $V$ while satisfying a safety constraint $\mathcal{C} = \{x : h(x) \geq 0\}$ can be obtained by solving the optimization problem 
\begin{equation}
    \label{eq:CLF-CBF-opt}
    \begin{array}{ll}
    \mbox{minimize} & u_{t}^{T}Ru_{t} \\
    \mbox{s.t.} & \Lambda^{d^{\prime}}_{i}(\hat{x}_{t,j})u_{t} \geq \overline{\omega}^{d^{\prime}}_{j} \ \forall j \in X^{d^{\prime}}_{t}(\delta) \quad \mbox{(HOSCBF)} \\
      & \Gamma_{i}(\hat{x}_{t,i})u_{t} \leq \overline{\tau}_{i} \ \forall i \in Y_{t}(\overline{V}) \quad \mbox{(CLF)}
      \end{array}
\end{equation}
at each time step, where $R$ is a positive definite matrix representing the cost of exerting control.

\subsection{HOSCBF-CLF Construction}
As in the case of a single HOSCBF constraint, satisfaction of (\ref{eq:CBF-CLF}) will depend on the geometry of the safe region and goal set as well as the values of $\gamma_{i}$ and $\theta_{ij}$. 
We consider a linear system with dynamics 
\begin{equation}
\label{eq:LTI-SDE}
dx_{t} = (Fx_{t} + Gu_{t}) \ dt + \sigma dW_{t}.   
\end{equation}
The goal set is ellipsoidal, so that $w(x) = V(x) = (x-x^{\prime\prime})^{T}\Psi(x-x^{\prime\prime})$, and the safe region $\mathcal{C}$ is given by a hyperplane constraint $a^{T}x-b \geq 0$.
We next construct SCBF-CLF as a special case of HOSCBF-CLF to ensure safety and stability of the cases where $\mbox{rank}(G) = n$ and $\mbox{rank}(G) < n$. 

\begin{proposition}
 \label{prop:CBF-CLF}
Suppose that $\mbox{rank}(G)=n$ and  the following conditions hold:
\begin{align}
    \label{eq:CBF-CLF-1}
    a^{T}x^{\prime\prime} - b >& 0 \\
    \label{eq:CBF-CLF-2}
    \left\{(x-x^{\prime\prime})^{T}\Psi(x-x^{\prime\prime}) \leq  \frac{\overline{\theta}^{2}\overline{\lambda}}{2}\right\} \cap 
    \{a^{T}x -  b \leq 0\} =& \emptyset 
\end{align}
 Then there exists $\delta > 0$ such that, at each time $t$, there exists $u$ satisfying (\ref{eq:CBF-CLF}) when $\overline{V} = \frac{\overline{\theta}^{2}\overline{\lambda}(\Phi)}{2}$.
\end{proposition}

\begin{proof}
Select $\delta$ such that $\delta < a^{T}x^{\prime\prime}$. We consider three cases. In the first case, $X_{t}(\delta) \neq \emptyset$ and $Y_{t}(\overline{V}) = \emptyset$. In the second case, $X_{t}(\delta) = \emptyset$ and $Y_{t}(\overline{V}) \neq \emptyset$. In the third case, $X_{t}(\delta) \neq \emptyset$ and $Y_{t}(\overline{V}) \neq \emptyset$.

If $X_{t}(\delta) \neq \emptyset$ and $Y_{t}(\overline{V}) = \emptyset$, then $u$ satisfying $a^{T}Gu > 0$ suffices to ensure safety by Lemma 1 in \cite{clark2020control}. 
If $Y_{t}(\overline{V}) \neq \emptyset$ and $X_{t}(\delta) = \emptyset$, then choose $u$ such that $Gu = -(\hat{x}_{t,i}-x^{\prime\prime})$ for some $i \in Y_{t}(\overline{V})$. 
By Proposition 1 in \cite{clark2020control}, for any positive definite matrix $\Phi$ and $x^{\prime} \in \mathbb{R}^{n}$ if $$
||\hat{x}_{t,i}-\hat{x}_{t,j}||_{2} \leq \overline{\theta} \leq \overline{\lambda}(\Phi)^{-1/2}\sqrt{2}
$$
and $\hat{x}_{t,i}$ and $\hat{x}_{t,j}$ both satisfy $(x-x^{\prime})^{T}\Phi(x-x^{\prime}) > (1-\epsilon)$ for $\epsilon$ sufficiently small, then $$(\hat{x}_{t,i}-x^{\prime})^{T}\Phi(\hat{x}_{t,j}-x^{\prime}) > 0.$$ Choosing $\Phi = \frac{1}{2\overline{\theta}^{2}}\Psi$ and $x^{\prime}=x^{\prime\prime}$ yields $\overline{\lambda}(\Phi) = \frac{1}{2\overline{\theta}^{2}}$, and hence $\overline{\theta} \leq \overline{\lambda}(\Phi)^{-1/2}\sqrt{2}$ holds by construction. Hence we have $$(\hat{x}_{t,i}-x^{\prime\prime})^{T}\left( \frac{1}{2\overline{\theta}^{2}\overline{\lambda}}\Psi\right)(\hat{x}_{t,j}-x^{\prime\prime}) > 0$$ when $\hat{x}_{t,i}$ and $\hat{x}_{t,j}$ satisfy $$(x-x^{\prime\prime})^{T}\left( \frac{1}{2\overline{\theta}^{2}\overline{\lambda}}\Psi\right)(x-x^{\prime\prime}) \geq 1,$$ or equivalently, when they satisfy $$(x-x^{\prime\prime})^{T}\Psi(x-x^{\prime\prime}) \geq \frac{\overline{\theta}^{2}\overline{\lambda}}{2}.$$



Finally, suppose that $Y_{t}(\overline{V}) \neq \emptyset$ and $X_{t}(\delta) \neq \emptyset$. Choosing $u$ such that $Gu = -(\hat{x}_{t,i}-x^{\prime\prime})$ for some $i \in Y_{t}(\overline{V})$. By the preceding discussion, (\ref{eq:CBF-CLF}) holds for all $j \in Y_{t}(\overline{V})$. By choice of $\delta$, $i \in X_{t}(\delta)$ implies that $a^{T}\hat{x}_{t,i} < a^{T}x^{\prime\prime}$. Hence, we have  $$a^{T}Gu = a^{T}(x^{\prime\prime}-\hat{x}_{t,i}) > 0,$$ and therefore $\Lambda(\hat{x}_{t,i})u_{t} < 0$ is satisfied for all $i \in X_{t}^{\prime}$. 
\end{proof}

We next turn to the case where $\mbox{rank}(G) < n$. As in the case of the SCBF construction in \cite{clark2020control}, we add a hyperplane constraint $(x-x^{\prime\prime})^{T}\Psi v < 0$ to ensure that the SCBF-CLF constraints are satisfied. 

\begin{proposition}
Suppose that $v \in \mbox{span}(G)$ and satisfies the following conditions: (i) $a^{T}v > 0$, and (ii) the initial state $x^{\prime}$ satisfies $(x^{\prime}-x^{\prime\prime})^{T}\Psi v < 0$. Then there exists $u$ satisfying (\ref{eq:CBF-CLF}) at each time $t$. 
\end{proposition}

\begin{proof}
At each time $t$, choose $Gu = v$. We need to verify both SCBF constraints and the CLF constraint for each $i$. First, for the constraint $h(x) = a^{T}x-b > 0$, we must have $-a^{T}Gu < 0$, which is equivalent to assumption (i) of the proposition. For the constraint $(x-x^{\prime\prime})^{T}\Psi Gu < 0$, the choice of $v = Gu$ and the hyperplane constraint $(x-x^{\prime\prime})^{T}\Psi v < 0$ implies that the CLF constraint is satisfied. Finally, the hyperplane constraint $(x-x^{\prime\prime})^{T}\Psi v < 0$ can be satisfied if $-v^{T}\Psi Gu < 0$, or equivalently, if $-v^{T}\Psi v < 0$, which holds since $\Psi$ is positive definite.
\end{proof}

%% file: sections/ActuatorFailure.tex
\section{Safe Control Under Actuator Failures}
\label{sec:PM_ActFail}
Motivated by FT-CBFs for sensor faults, in this section, we propose a passive fault-tolerant control based on CBFs to mitigate actuator failures. 
We consider a nonlinear control system affected by actuator failures in a noise-free scenario. The system dynamics can be described as 
\begin{eqnarray}
\label{eq:afstate-sde}
dx_{t} &=& (f(x_{t})+g(x_{t})u^F_{t}) \ dt \\
\label{eq:afoutput-sde}
dy_{t} &=& cx_{t}  \ dt,
\end{eqnarray}
where $u^F_t$ represents the actuator's output with failures. 

We consider loss of control effectiveness failure \cite{ye2006adaptive} in which a subset of actuators produce a zero output. Let $\mathcal{I}$ denote the set of failed actuators. The actuator failure model is given by $u^F_t = L u_t$, where $L$ is a $p \times p$ diagonal matrix with $L_{ii}=0$ if $i\in \mathcal{I}$ and $L_{ii}=1$ otherwise. 


The concept of actuator redundancy for LTI systems is proposed in~\cite{zhao1998reliable}. We extend the definition to nonlinear systems. 
\begin{Definition}[Actuator Redundancy]
\label{def:redundancy}
A nonlinear system \eqref{eq:afstate-sde}-\eqref{eq:afoutput-sde} is said to have $r$ actuator redundancy if the system remains controllable 
for all failure patterns  $L_j \in \mathcal{L}_r=\{ L_\mathcal{I} \mid \mathcal{I}\subsetneq \{1,\ldots p\}, |\mathcal{I}|\leq r\}$, where $|\mathcal{I}|$ denotes the cardinality of $\mathcal{I}$.
\end{Definition}

In this section, we assume that the system described in \eqref{eq:afstate-sde} and \eqref{eq:afoutput-sde} has $r$ actuator redundancy.

\noindent\textbf{Problem Statement:} Given a set $\mathcal{C}$ defined in \eqref{eq:safe-region}, construct a control policy that, at each time $t$, maps the sequence $\{y_{t^{\prime}} : t^{\prime} \in [0,t)\}$ to an input $u_{t}$ and, for any failure $\mathcal{L}_j\in\mathcal{L}$, ensure $x_t \in \mathcal{C}$, $\forall t$. 

The goal is to ensure the safety defined in \eqref{eq:safe-region} of the system when actuator failures occur. 
The intuition behind our approach is to choose a control input $u$ that is safe for all possible actuator failure patterns. To achieve this, we examine the system dynamics with $m$ possible failure patterns $\mathcal{L}_j\in\mathcal{L}$ at each time and choose $u_t$ to satisfy all safety constraints. 

\begin{Lemma}
\label{lemma:afCBF}
For a system (\ref{eq:afstate-sde})--(\ref{eq:afoutput-sde}) with safety region defined by (\ref{eq:safe-region}), $u_{t}$ is chosen to satisfy constraints $\bigcap_{\mathcal{L}_j\in\mathcal{L}}\Omega_j$ defined as follows.
\begin{equation}
    \label{eq:afCBFs}
    \Omega_j = \{u : \frac{\partial h}{\partial x}\left(
    f(x_t)+g(x_t) L_j u_t\right) \geq -\alpha(h(x_t))\}
\end{equation}
Then when the set $\bigcap_{L_j\in\mathcal{L}}\Omega_j$ is non-empty, the safety can be guaranteed when any failure pattern $L_j\in\mathcal{L}$ happens. 
\end{Lemma}
\begin{proof}
By Corollary 2 in \cite{ames2016control}, safe set $\mathcal{C}$ is forward invariant if $u$ satisfies the CBF constraint. Safety of the system under actuator failure $i$ can be ensured by choosing $u\in \bigcap_{\mathcal{L}_j\in\mathcal{L}}\Omega_j \subseteq \Omega_i$. 
\end{proof}

However, a feasible control input $u$ may not exist when $\frac{\partial h}{\partial x}g(x)L_j=0$, since $u$ does not affect states $x$ due to dynamics or actuator failures. To address this problem, we define HOCBFs for actuator failure such that for $d$-th degree $\frac{\partial h^d}{\partial x}g(x)L_j\neq0$. Then, we choose control input $u$ to satisfy constraints constructed by HOCBFs.

\begin{Lemma}
\label{lemma:afHOCBF}
For a system (\ref{eq:afstate-sde})--(\ref{eq:afoutput-sde}) with safety region defined by (\ref{eq:safe-region}), suppose there exist $m$ relative degrees $d^{\prime}_j$, for each $L_j,\ j\in\{1,\ldots,m\}$ such that $\frac{\partial h^{d^{\prime}_j}}{\partial x}g(x_t) L_j\neq 0$. For all $x_0\in\overline{\mathcal{C}}:= \cap_{j=1}^m\cap_{d=0}^{d’_j} \mathcal{C}_j^d$, $u_{t}$ is chosen to satisfy constraints $\bigcap_{L_j\in\mathcal{L}}\Omega_j$ defined as follows.
\begin{equation}
    \label{eq:afHOCBFs}
    \Omega_j = \{u:\frac{\partial h^{d^{\prime}_j}}{\partial x}\left(
    f(x_t)+g(x_t) L_j u\right) \geq -\alpha(h^{d^{\prime}_j}(x_t))\}
\end{equation}
Then when the set $\bigcap_{L_j\in\mathcal{L}}\Omega_j$ is non-empty, the safety can be guaranteed when any failure pattern $L_j\in\mathcal{L}$ happens. 
\end{Lemma}
\begin{proof}
For a given $j\in\{1,\ldots,m\}$, the safe set $\cap_{d=0}^{d’_j} \mathcal{C}_j^d$ remains forward invariant by Theorem \ref{th:hocbf}, if $u_t$ is chosen to satisfy the corresponding HOCBF constraint \eqref{eq:afHOCBFs}. For an unknown actuator failure $i$, the forward invariance of the set $\overline{\mathcal{C}}= \cap_{j=1}^m\cap_{d=0}^{d’_j} \mathcal{C}_j^d$ is ensured if $u_t\in \bigcap_{\mathcal{L}_j\in\mathcal{L}}\Omega_j \subseteq \Omega_i$ for $x_0\in\overline{\mathcal{C}}$.
\end{proof}

Although the approach in Lemma \ref{lemma:afCBF} and Lemma \ref{lemma:afHOCBF} can ensure the safety of the system, excessive constraints make the existence of a feasible solution problematic. In what follows, we present feasibility verification for safe control under actuator failures. 

\subsection{Feasibility Verification}

We provide feasibility verification for both CBF and HOCBF of the system with actuator failures. 

\subsubsection{Verification for CBF Under Actuator Failures}
We first show the verification for CBF of the system with actuator failures. We denote $A(x)$ and $\Xi(x)$ as follows
\begin{equation*}
    \begin{split}
        A(x) &= \left( -\frac{\partial h}{\partial x} g(x)L_1, \ldots, -\frac{\partial h}{\partial x} g(x)L_p \right)^{T}, \\
        \Xi(x) &= [\xi(x), \ldots, \xi(x)]^T,
    \end{split}
\end{equation*}
where 
\begin{equation*}
    \xi(x) = \frac{\partial h}{\partial x}f(x_t) + h(x_t)
\end{equation*}

\begin{proposition}
\label{prop:verify_ftcbf}
There exists a feasible solution $u$ satisfying a set of $m$ CBF constraints if and only if there is no $x$ and $y$ satisfying $A^{T}(x)y=0$, $y\geq 0$ and $\Xi^{T}(x)y<0$.
\end{proposition}
\begin{proof}
By Corollary \ref{coro:Farkas}, we have a solution $u\in\mathbb{R}^{p}$, if and only if there does not exist $y\in\mathbb{R}^{m}$ such that \eqref{eq:coro1_y} hold. 
Conversely, if for some $x_0$ and $y_0$ satisfying $\frac{\partial h}{\partial x} g(x_0) L_j = 0$ for some $j$,  $A^{T}(x)y_0=0$, $y_0\geq 0$ and $\Xi^T(x)y_0<0$, the set $\mathcal{C}$ is not positive invariant. 
\end{proof}

Based on this proposition, we can formulate the following conditions via the Positivstellensatz.

\begin{Lemma}
\label{Lemma:verify_ftcbf}
There exists a feasible solution $u$ satisfying a set of $m$ CBF constraints 
if and only if there exist polynomials $\rho^0_i(x,y)$, $\rho^1_i(x,y)$, sum-of-squares polynomials $q_S(x,y)$, and integers $r$ such that

\begin{equation}
    \label{eq:verify_afcbf}
    \phi(x,y) + \chi(x,y) + \psi(x,y) = 0, 
\end{equation}
and
\begin{align*}
    \phi(x,y) =& \sum_{S\subseteq \{1,\ldots,3\}} q_{S}(x,y)\prod_{i\in S}\phi_i(x,y)  \\
    \chi(x,y) =& \left(\Xi^{T}y\right)^{2 r}\\
    \psi(x,y)  =& \sum_{j=1}^{m}\left( \sum_{i=1}^{m}\rho^0_i(x,y)\left[A^T y\right]_{i} \right),
\end{align*}
where $\phi_1(\cdot)=y$, $\phi_2(\cdot)=-\Xi^T(x)y$ and $\phi_3(\cdot)=h(x)$. 
\end{Lemma}
\begin{proof}
By Proposition \ref{prop:verify_ftcbf}, we have $h(x)$ is an CBF if and only if there exist no  $x$ and $y$ satisfying $y_j\geq 0$, $\frac{\partial h}{\partial x} g(x) L_j = \boldsymbol{0}\ \forall j\in\{1,\ldots,m\} $,  $A^{T}y=\boldsymbol{0}$ and $\Xi^{T}y<0$. 
The conditions are equivalent to $\forall j\in\{1,\ldots,m\}$,
\begin{equation*}
    h(x)\geq 0, \ A^{T}y=\boldsymbol{0}, \ y_j\geq 0, \ -\Xi^{T}(x)y\geq 0, \Xi^{T}(x)y\neq 0
\end{equation*}
These conditions are equivalent to \eqref{eq:verify_afcbf} by the Positivstellensatz.
\end{proof}

\subsubsection{Verification for HOCBF of Actuator Failures}

We next verify the feasibility of a set of HOCBF constraints. We denote $A$ and $\Xi$ as follows
\begin{equation*}
    \begin{split}
        A(x) &= \left( -\frac{\partial h^{d^{\prime}_1}}{\partial x} g(x)L_1, \ldots, -\frac{\partial h^{d^{\prime}_p}}{\partial x} g(x)L_p \right)^{T}, \\ 
        \Xi(x) &= [\xi_1(x), \ldots, \xi_p(x)],
    \end{split}
\end{equation*}
where 
\begin{equation*}
    \xi_j(x) = \frac{\partial h^{d^{\prime}_j}}{\partial x}f(x_t) + h(x_t)
\end{equation*}

\begin{proposition}
\label{prop:verify_aftcbf}
There exists a feasible solution $u$ satisfying a set of $m$ CBF constraints if and only if there is no $x$ and $y$ satisfying $A^{T}y=0$, $y_j\geq 0$ and $\Xi^{T}(x)y<0$, $\forall j\in\{1,\ldots,m\}$. 
\end{proposition}
\begin{proof}
By Definition \ref{Def:HOCBF}, we have $h^{d}(x_j)\geq 0$ for all $x \in \overline{\mathcal{C}}$, if and only if the following two conditions are satisfied. For all $x \in \overline{\mathcal{C}}$, $h^{d}(x)\geq0$ for all $d\leq d^{\prime}$. Moreover, $\frac{\partial h^{d^{\prime}}}{\partial x}g(x)L_j \neq 0$ and $u$ are chosen to satisfy \eqref{eq:HOCBF} for all $j$ at the boundary. By Corollary \ref{coro:Farkas}, we have a solution $u\in\mathbb{R}^{p}$, if and only if there does not exist $y\in\mathbb{R}^{m}$ such that \eqref{eq:coro1_y} holds. 
Conversely, if for some $x_0$ and $y_0$ satisfying   $A^{T}y_0=0$, $y_0\geq 0$ and $\Xi^{T}(x)y_0<0$, the set $\mathcal{C}$ is not positive invariant. 
\end{proof}

Based on this proposition, we can formulate the following conditions via the Positivstellensatz.

\begin{Lemma}
\label{Lemma:verify_afhocbf}
There exists a feasible solution $u$ satisfying a set of $m$ HOCBF constraints 
if and only if there exist polynomials $\rho^0_i(x,y)$, $\rho^1_i(x,y)$, sum-of-squares polynomials $q_S(x,y)$, integers $s=2+\sum_{j=1}^{m}d^{\prime}_j$, and $r_1,\ldots,r_m$ such that
\begin{equation}
    \label{eq:verify_afhocbf}
    \phi(x,y) + \chi(x,y) + \psi(x,y) = 0, 
\end{equation}
and
\begin{align*}
    \phi(x,y) =& \sum_{S\subseteq \{1,\ldots,s\}} q_{S}(x,y)\prod_{i\in S}\phi_i(x,y)  \\
    \chi(x,y) =& \prod_{\forall j\in\{1,\ldots,m\}}\left(-\Xi^{T}(\hat{x}_j)y\right)^{2 r_j}\\
    \psi(x,y)  =& \sum_{j=1}^{m}\left( \sum_{i=1}^{m}\rho^0_i(x,y)\left[A^T y\right]_{i} \right),
\end{align*}
where $\phi_1(\cdot)=y$, $\phi_2(\cdot)=-\Xi^T(x)y$ and for $d_j\in\{0,\ldots,d^{\prime}_j\}$ $\phi_{\{3,2+\sum_{j=1}^{m}d^{\prime}_j\}}(\cdot)=\hat{h}^{d_j}(\hat{x}_j)$.
\end{Lemma}
\begin{proof}
By Proposition \ref{prop:verify_ftscbf}, we have $h(x)$ is an HOCBF if and only if there exist no  $x$ and $y$ such that $h^{d^{\prime}_j}(x)\geq \boldsymbol{0}$, for all $x\in \overline{\mathcal{C}}$, $y_j\geq 0$, $A^{T}y=\boldsymbol{0}$, and $\Xi^{T}(x)y<0$. 
The conditions are equivalent to $\forall j\in\{1,\ldots,m\}$, 
\begin{equation*}
    \begin{split}
    &
    h^{d_j}(x) \geq 0, \forall d_j\in\{1,\ldots,d_j^{\prime}\} \\
    &A^{T}(x)y=\boldsymbol{0}, \ y\geq 0, \ -\Xi^{T}(x)y\geq 0, \Xi^{T}(x)y\neq 0
    \end{split}
\end{equation*}
These conditions equal to \eqref{eq:verify_afhocbf} by the Positivstellensatz.
\end{proof}

%% file: sections/CaseStudy.tex
\section{Case Study}
\label{sec:simulation}
\begin{figure*}[tp!]
\centering
\begin{subfigure}{.4\textwidth}
    \includegraphics[width=\textwidth]{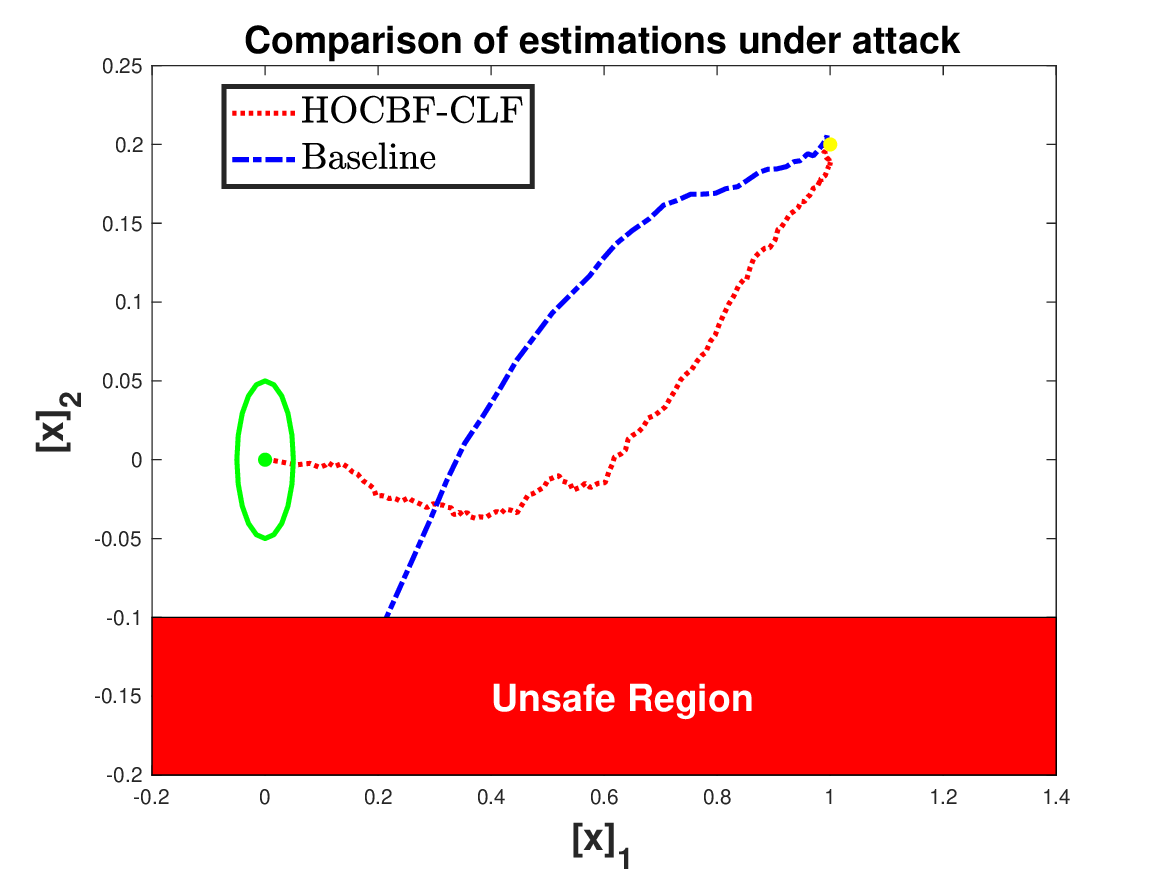}
    \caption{}
    \label{fig:trj_hoscbf-clf}
\end{subfigure}
\begin{subfigure}{.4\textwidth}
    \includegraphics[width=\textwidth]{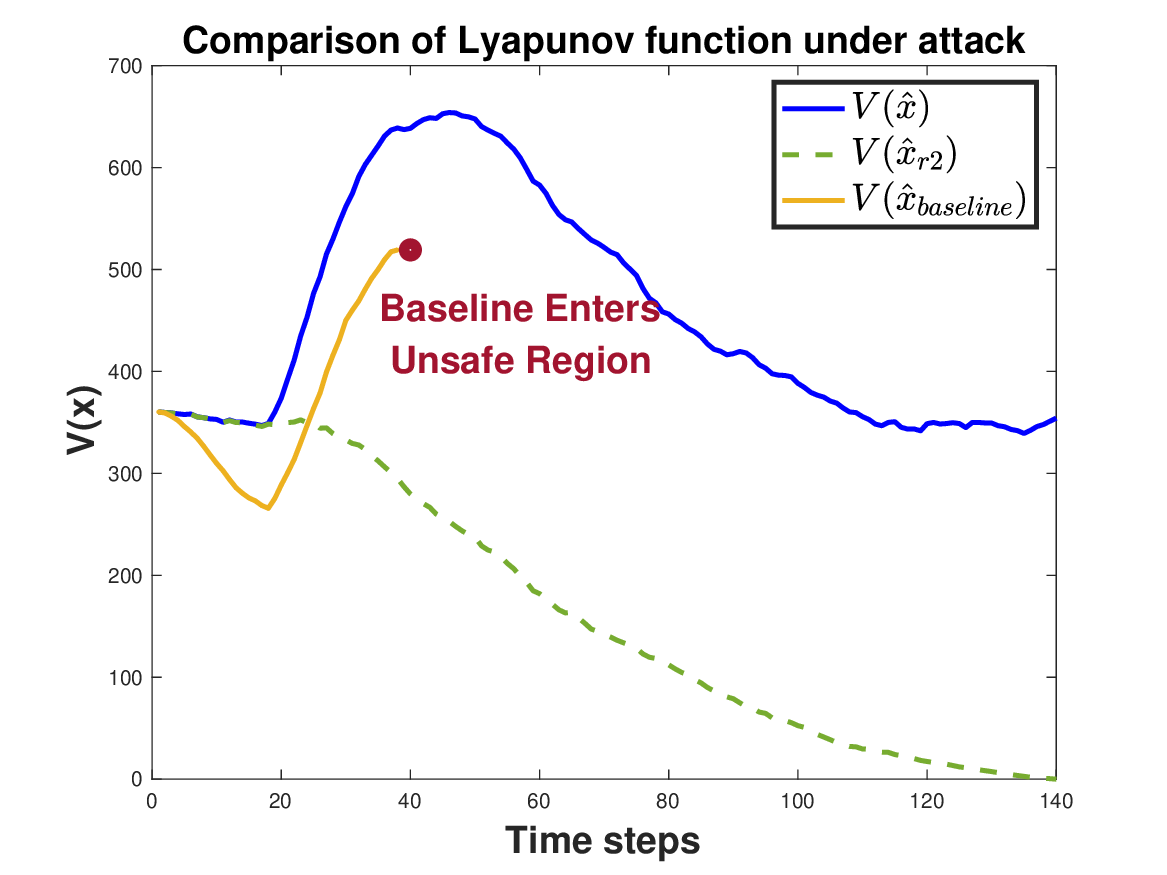}
    \caption{}
    \label{fig:vf_hoscbf-clf}
\end{subfigure}
\caption{Comparison of actual trajectory and Lyapunov function between HOSCBF-CLF and baseline on WMR system under sensor false data injection attacks. In (\subref{fig:trj_hoscbf-clf}), the baseline entered unsafe region while proposed method remains safe and converge to goal region. In (\subref{fig:vf_hoscbf-clf}), the Lyapunov function of real states decreases and converges to zero. }
\label{fig:sim_sf}
\end{figure*}

In this section, we present a case study of a wheeled mobile robot under sensor faults and a case study of a Boeing 747 under actuator Failure. We first describe the system models then then present the results. 
\subsection{Sensor Fault and Attack}
\label{subsec:system_model}
We consider a wheeled mobile robot (WMR) with dynamics
\begin{equation}
\begin{pmatrix}[\dot{x}_t]_1\\ [\dot{x}_t]_2\\ \dot{\theta}_t\end{pmatrix}
=\begin{pmatrix}\cos{\theta_t}&0\\\sin{\theta_t}&0\\0&1\end{pmatrix}\begin{pmatrix}[\omega_t]_1\\ [\omega_t]_2\end{pmatrix}
+\mathbf{w}_t
\label{original_model}
\end{equation}
where $([x_t]_1, [x_t]_2, \theta_t)^T$ is the vector of the horizontal, vertical, and orientation coordinates for the wheeled mobile robot, $([\omega_t]_1, [\omega_t]_2)^T$ (the linear velocity of the robot and the angular velocity around the vertical axis) is taken as the control input, and $\mathbf{w}_t$ is the process noise. 
The feedback linearization \cite{chen2018building} is utilized to transform the original state vector and the WMR model into the new state variable $x_t = ([x_t]_1, [x_t]_2, [\dot{x}_t]_1, [\dot{x}_t]_2)^T$ with control input $u_t = ([u_t]_1, [u_t]_2)^T$ and the controllable linearized model defined as follow. 
\begin{equation}
\label{linearized_model}
\dot{x}_t = F x_t + G u_t+\mathbf{w}_{t}^{\prime}
\end{equation}
where the process noise $\mathbf{w}_{t}^{\prime}\in \mathbb{R}^{4}$ has distribution $\mathcal{N}(0,\sigma_{w} I)$, where $\sigma_{w}=0.05$. The matrices $F$ and $G$ are defined as 
\begin{equation*}
    F = \begin{pmatrix}0&0&1&0\\0&0&0&1\\0&0&0&0\\0&0&0&0\end{pmatrix},\ 
    G = \begin{pmatrix}0&0\\0&0\\1&0\\0&1\end{pmatrix}. 
\end{equation*}
The following compensator is used to calculate the input $[\omega_t]_1$ and $[\omega_t]_2$ into (\ref{original_model})
\begin{eqnarray}
\label{compensator1}
[\omega_t]_1 &=& \int_{t^-}^{t^+} [u_t]_1\cos{\theta_t} + [u_t]_2\sin{\theta_t} \ dt \\
\label{compensator2}
[\omega_t]_2 &=& ([u_t]_2\cos{\theta_t} - [u_t]_1\sin{\theta_t})/[\omega_t]_1.
\end{eqnarray}
Here we assume that the observation for the orientation coordinate $\theta_t$ is attack-free and noise-free, which enables feedback linearization based on the variable $\theta_{t}$. 

In the linearized model, we use the observation equation
\begin{equation}
\begin{pmatrix}[y_t]_1\\ [y_t]_2\\ [y_t]_3\\ [y_t]_4\\ [y_t]_5\\ [y_t]_6\end{pmatrix}
=\begin{pmatrix}1&0&0&0\\1&0&0&0\\0&1&0&0\\0&1&0&0\\0&0&1&0\\0&0&0&1\end{pmatrix} \begin{pmatrix}[x_t]_1\\ [x_t]_2\\ [\dot{x}_t]_1\\ [\dot{x}_t]_2\end{pmatrix}+\mathbf{a}_t
+\mathbf{v}_t
\label{obervation_simulation}
\end{equation}
where the measurement noise $\mathbf{v}_t\in \mathbb{R}^{6}$ has distribution $\mathcal{N}(0,\sigma_{v} I)$, where $\sigma_{v}=0.05$. The impact of the attack is denoted as $\mathbf{a}_t$. The attack signal satisfies that
$$\mathbf{a}_t =  \begin{cases}
    \mathbf{0}, &\ t<1 \\
    [0,0,0,2,0,0]^{T}, &\ t\geq 1
\end{cases}$$
Note that there is one redundant sensor for the horizontal coordinate and one for the vertical coordinate.

Here we let the safe region $\mathcal{C} = \{x_t : h(x_t) = [x_t]_2 + 0.1 \geq 0, t \geq 0\}$ and the goal region $\mathcal{G}  = \{x_t : \omega(x_t)=d-||x_t - x_g||_2 \geq 0\}$, where $x_g$ is $(0,0)$ and $d = 0.05$ is the radius of the goal region. 
The baseline utilizes a fault detection scheme~\cite[Chapter 7.3]{blanke2006diagnosis} to detect and identify sensor faults by comparing EKF residuals against the threshold $0.1$ and recomputes control input with an LQR controller. We then compare with our proposed HOCBF-based and CLF-based method. 
To keep the system remaining within safe region, we systematically construct the FT-SCBF with relative degree $1$ by using the following class of sets 
\begin{equation*}
    \begin{split}
        C^{(0)} = \{x\mid h^0(x_t) &= a^T x_t + b \geq0, \forall t\geq 0\}\\
        C^{(1)} = \{x\mid h^1(x_t) &= a^T F x_t + a^T x_t +b \geq 0, \forall t\geq 0\},
    \end{split}
\end{equation*}
where $a^T=[0,1,0,0]$ and $b^T=[0,0.1,0,0]$. This differs from our previous work \cite{clark2020control} which solves the problem by manually tuning the parameters and constructing the CBFs for high relative degree.  
In order to reach the goal region without violating the safety constraint, we choose the CLF
\begin{equation}
V(x) = (x_t - x_g)^TP_d(x_t - x_g)
\label{CLF}
\end{equation}
where $P_d = \begin{pmatrix}\frac{1}{d}I&0\\0&I\end{pmatrix}P_L\begin{pmatrix}\frac{1}{d}I&0\\0&I\end{pmatrix}$, $P_L$ is the solution of the Lyapunov equation $F^{T}P_{L} + P_{L}F = -I$, and $I$ is the identity matrix  \cite{nguyen2016exponential}, \cite{ames2014rapidly}. We set $\rho = 0.2$, $\eta=0.8$ and $M=2$ in the CLF constraint. The control input $u_t$ is computed at each time step by solving (\ref{eq:CLF-CBF-opt}) with $R = I$.



\textbf{Simulation Result: }
The results are shown in Fig.~\ref{fig:sim_sf}. In Fig.~\ref{fig:sim_sf}(a), we plot the first two dimensions of the state, which describe the horizontal and vertical coordinates. Note that the robot stays in the safe region and eventually reaches the goal region, and hence satisfies  safety and stability. As a comparison, the baseline can identify sensor faults but still resulted in a safety violation due to the slow response time of residual-based diagnosis.

\subsection{Actuator Failure}
In lateral control of an aircraft, lower yaw rate can renders smoother flight performance to avoid package damage or harsh passenger experience. 
We consider the lateral dynamics of Boeing 747 with state $x(t)=\left[[x]_1,[x]_2,[x]_3,[x]_4\right]^{T}$, where $[x]_1$ is the side-slip angle, $[x]_2$ is the yaw rate, $[x]_3$ is the roll rate, $[x]_4$ is the roll angle. 
In this case study we study yaw rate control with preset upper and lower boundary on yaw rate $\mathcal{C} = \{x_t : -0.025\leq [x_t]_2 \leq 0.025, t \geq 0\}$ and reference point $x_t^T=[0,0,0,0]$. 
The dynamic system can be linearized and described as follows. 
\begin{equation}
    \begin{split}
        \dot{x}_t &= F x_t + G u^{F}_{t}+\mathbf{w}_{t}\\
        y_t &= \left[\begin{array}{cccc}
            0 & 1 & 0 & 0
        \end{array}\right]x_t + \mathbf{v}_{t},
    \end{split}
\end{equation}
where the process noise $\mathbf{w}_{t}\in \mathbb{R}^{4}$ has distribution $\mathcal{N}(0,\sigma_{w} I)$, where $\sigma_{w}=0.001$ and the measurement noise $\mathbf{v}_{t}\in \mathbb{R}$ has distribution $\mathcal{N}(0,\sigma_{v} I)$, where $\sigma_{v}=0.001$. The matrices $F$, $G$ are defined as 
\begin{equation*}
\begin{split}
    F&=\left[\begin{array}{cccc}
    -0.0558 & -0.9968 & 0.0802 & 0.0415 \\
    0.598 & -0.115 & -0.0318 & 0 \\
    -3.05 & 0.388 & -0.465 & 0 \\
    0 & 0.0805 & 1 & 0
    \end{array}\right],\\
    G&=\left[\begin{array}{ccc}
    0.00729 & 0.01 & 0.005 \\
    -0.475  & -0.5 & -0.3 \\
    0.153  & 0.2  & 0.1 \\
    0 & 0 & 0
    \end{array}\right].
\end{split}
\end{equation*}

We assume that the system has two redundant actuators. The control input $u^{F}_t$ contains three control signals representing three rudder servos, which may fail and output zero when failure happens. 
We simulate the actuator failure by denoting $u^{F}_t = L u_t$ with two potential failure patterns $L_{1} = diag([1,0,1])$ and $L_{2} = diag([0,1,1])$. Failure $L_{1}$ starts from $2.5s$ to $5s$ and failure $L_{2}$ occurs since $5s$. We set $\rho = 0.2$, $\eta=0.8$ and $M=2$ in the CLF constraint. 

The baseline utilizes a fault detection scheme~\cite[Chapter 7.3]{blanke2006diagnosis} to detect and identify actuator failures by comparing the EKF residuals against the threshold $0.02$. Once identifying actuator failure $L_i$, the baseline recomputes control input based on reconfigured $G=GL_i$ by solving a CBF-CLF-based quadratic program.
In order to keep the system remaining within safe region, we further define the following class of set for the upper and lower bound of yaw rate as
\begin{equation*}
    \begin{split}
        C_0 = \{x\mid h_0(x_t) &= a_0^T x_t + b_0 \geq0, \forall t\geq 0\}\\
        C_1 = \{x\mid h_1(x_t) &= a_1^T x_t + b_1 \geq 0, \forall t\geq 0\},
    \end{split}
\end{equation*}
where $a_0^T=[0,1,0,0]$, $a_1^T=[0,-1,0,0]$ and $b_0^T=b_1^T=[0,0.025,0,0]$. 
We impose CBF with relative degree $0$ as constraints to ensure safety. 


\begin{figure}[tp!]
\centering
    \includegraphics[width=0.4\textwidth]{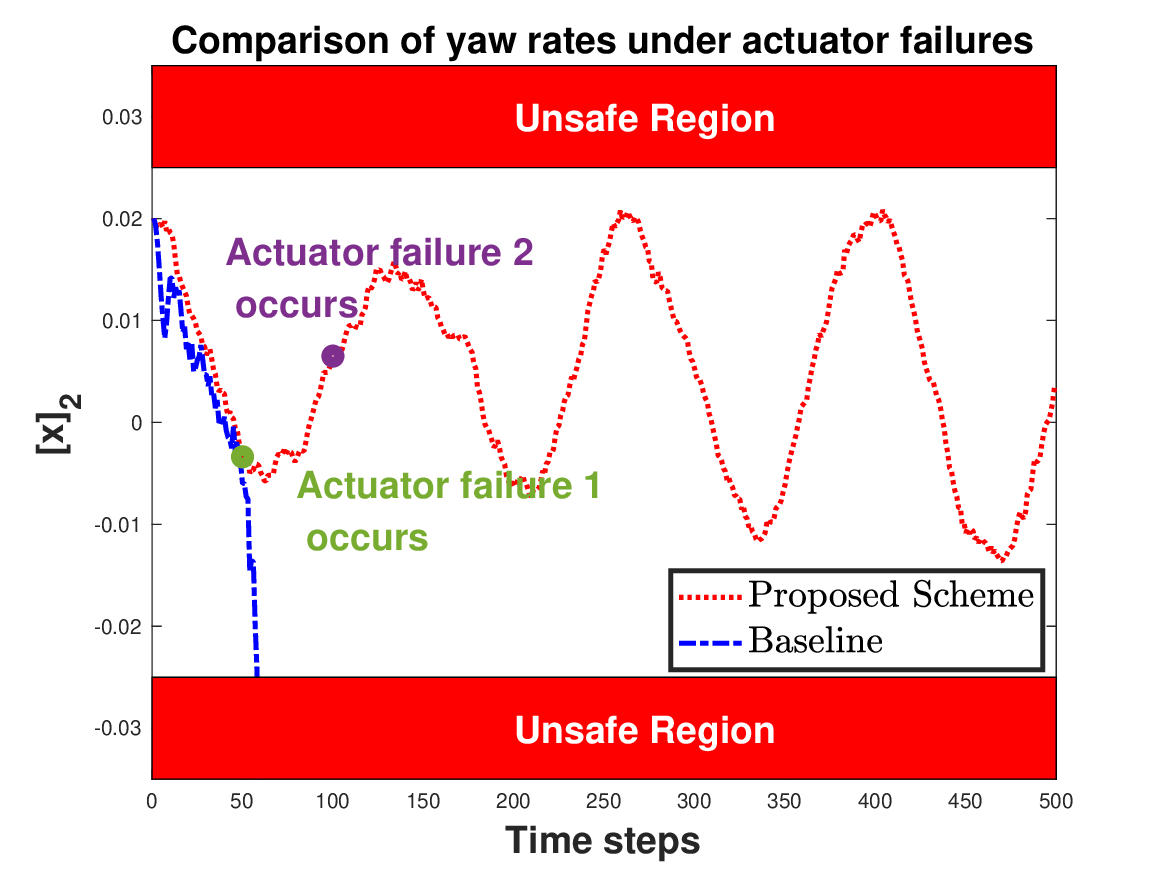}
    \caption{Yaw rate comparison between FT-HOCBF and baseline on Boeing 747 lateral control system with rudder servo failures. The baseline enters unsafe region when actuator failure 1 happens, while proposed method remains safe under either actuator failure 1 and 2. }
    \label{fig:yawrate}
\end{figure}

\textbf{Simulation Result: }
As is shown in Fig.~\ref{fig:yawrate}, we compare the yaw rate trajectory between the baseline and the proposed safe control scheme. The baseline identifies actuator failures but still results in safety violations. The trajectory of the proposed scheme stays in the safe region and converges to $0$, and hence satisfies safety and stability. 